\documentclass[aps,pra,groupedaddress,twocolumn, amsmath]{revtex4-2}

\usepackage{graphicx}
\usepackage{dcolumn}
\usepackage{bm}
\usepackage{lineno}

\usepackage{xcolor} 

\usepackage{xfrac}
\usepackage{hyperref}
\usepackage{comment}
\usepackage{xcolor}
\usepackage{upgreek}

\begin{document}

\title{Photo-birefringent effects in crystalline AlGaAs mirror coatings}


\author{C. Y. Ma, J. Yu, T. Legero, S. Herbers, D. Nicolodi, M. Kempkes, F. Riehle and U. Sterr}


\affiliation{Physikalisch-Technische Bundesanstalt, Bundesallee 100, 38116 Braunschweig, Germany}


\date{\today}

\begin{abstract}
High-reflective crystalline GaAs/Al$_{0.92}$Ga$_{0.08}$As coatings show reduced Brownian noise compared to conventional dielectric coatings. 
However, several ultra stable laser systems observed additional noise sources that  hinder the realization of the expected improvements in frequency stability. These additional noise sources are related to the birefringence of the coatings which can also be modified by intracavity light. While the origin of the birefringence is not yet well understood, its modification via illumination remains also unexplained.
Here we present an extensive study on the steady-state and transient modification of the birefringence by intracavity light and by uniform illumination at various wavelengths using an optical cavity at room temperature.
We find a unified description that suggests a primary two-photon process for photon energies below the bandgap of GaAs, or a single-photon process at higher energies.
Adding external illumination allows us to reduce noise induced by laser power fluctuations by balancing the photo-thermo-optic response of the mirrors and the photo-birefringent effect at more favorable low intracavity power levels.   
\end{abstract}


\maketitle

\section{Introduction}
Ultra-high reflectivity mirror coatings are indispensable for Fabry-Perot cavities required for optical atomic clocks \cite{lud15} or for gravitational wave detectors \cite{har06b}.
In these state-of-the-art instruments, the frequency or length stability is fundamentally limited by Brownian thermal noise of the mirror coatings \cite{har06b, gra20a}.
The Brownian thermal noise can be reduced using materials with smaller mechanical loss: crystalline AlGaAs coatings exhibit 
Brownian thermal noise power spectral density 16 times smaller than conventional dielectric coatings. 
This has been confirmed in both room-temperature and cryogenic cavities \cite{col16,yu23a}. However, ultra stable cavities with these coatings showed additional noise contributions \cite{yu23a,ked23} related to the so far not well understood birefringence of the coatings.
In ultra stable cavities operating at 1070\,nm, 1397\,nm and 1542\,nm, it has been observed that the birefringence of the coatings can be modified by intracavity light \cite{ked23,yu23a,kra25a,zhu23,ma24a}.  
It was also observed that the coating birefringence can be modified with additional illumination \cite{ma24a,wu25c}.
In reference \cite{ma24a} a 500\,Hz shift was observed at an intracavity power level of 1\,W for 1542\,nm or at about 100\,nW for diffuse light at 450\,nm reaching the area of the coating. 

Understanding the underlying mechanism of these effects might help to optimize next generation of crystalline coatings. 
%
In this work we employ a room-temperature ultra low expansion glass (ULE) cavity with fused silica substrates and crystalline coatings. 
We investigate the responses of the coating birefringence to intracavity radiation at 1542\,nm and to diffuse radiation at wavelengths between 450\,nm and 890\,nm.
We can describe the modification of the birefringence with a single effect driven by a one or two photon absorption process depending on the wavelength.
To study the dynamics of this effect we investigated the temporal response of the birefringence to step change of intracavity power. We offer an empirical model that takes into account the effect of intracavity power and additional uniform illumination.

The sensitivity of the coating birefringence to intracavity power couples the cavity resonance frequencies to laser power fluctuations and hence induces frequency noise. 
Power fluctuations also change the frequency by thermal effects in the substrate and coating (photo-thermo-optic effect), which is significant for fused silica substrates commonly employed in room temperature cavities.
It has been demonstrated that by choosing a suitable polarization, both effects could be largely canceled at a certain intracavity power \cite{kra25a, zhu23}.
Our model well predicts the response with additional LED light. 
This allows to cancel both effects at low intracavity power where the absolute intracavity power fluctuations are smaller and thus the related frequency stability is improved.

\section{Experimental setup}

We study AlGaAs coatings with a room temperature resonator operating at 1542\,nm, which is a copy of a previous cavity using dielectric coatings at 698\,nm \cite{hae15a}.
This design employs a 48\,cm long ULE spacer and a pair of 25.4\,mm diameter fused silica substrates (one plane and one concave with 1.0\,m radius of curvature). 
AlGaAs coatings with 8\,mm diameter \cite{col23} are bonded onto the substrates.   
They consist of 38.5 pairs of alternating Al$_{0.92}$Ga$_{0.08}$As/GaAs layers, of thickness 133.2\,nm and  115.6\,nm respectively, starting and ending with a GaAs layer. 
The $1/e^2$ beam radii at the plane mirror and at the concave mirror are 496\,$\upmu$m and 687\,$\upmu$m respectively. 
ULE rings are attached to the back of the mirrors to compensate for the coefficient of thermal expansion (CTE) mismatch between ULE and fused silica \cite{leg10}, leading to a measured CTE zero crossing at 297\,K \cite{ma24a}. 
Similar to other optical resonators with AlGaAs coatings \cite{ked23,yu23a,kra23a,zhu23}, we align the birefringent axes of the mirrors, 
hence maximizing the frequency splitting $\Delta_\mathrm{biref}$ between the two polarization eigenmodes (fast and slow) of 104\,kHz. 
This corresponds to a birefringence $\Delta n_\mathrm{biref}$ = $\frac{\Delta_\mathrm{biref}}{2\nu}$$\frac{L_\mathrm{cav}}{l_\mathrm{pen}}$ $\approx 4.5 \times 10^{-4}$,
where $\Delta_\mathrm{biref}$ is the frequency splitting between the two axes, $\nu$ = 194.43\,THz is the operating frequency of the resonator, $L_\mathrm{cav}$ is the optical length of the cavity, $l_\mathrm{pen}$ = 286\,nm is the
penetration depth of the light field in the optical coatings and the factor 2 accounts for the pair of mirrors in the resonator \cite{yu23}. 
%
%
Two lasers are locked independently from opposite sides of the cavity each to a fundamental Hermite–Gaussian (HG$_{00}$) mode \cite{yu23a, ma24a} that are separated by one free spectral range of 312\,MHz.
The cavity finesse of an HG$_{00}$ mode is $1.290 (2) \times 10^{5}$ and $1.282 (2) \times 10^{5}$ for slow and fast axis, respectively.
The frequency difference between adjacent cavity modes is largely immune to fluctuations of the cavity resonant frequency. 

The beat signal between the two lasers is detected from the transmitted and reflected beam at one side of the cavity. This common-path geometry avoids influence from path length fluctuations. 
This facilitates the study of coating specific photo effects that modify their birefringence. 

The average frequency of the two lasers can be calculated from an additional optical beat with another single crystalline silicon cavity operated at 124\,K with a frequency stability at $4 \times 10^{-17}$ \cite{mat17a}. 
The average of the two frequencies removes anti-correlated frequency changes due to mirror birefringence on the mirrors. 
This enables investigations of polarization-independent effects such as the photo-thermo-optic effect. 

\begin{figure}
	\includegraphics[width=\columnwidth]{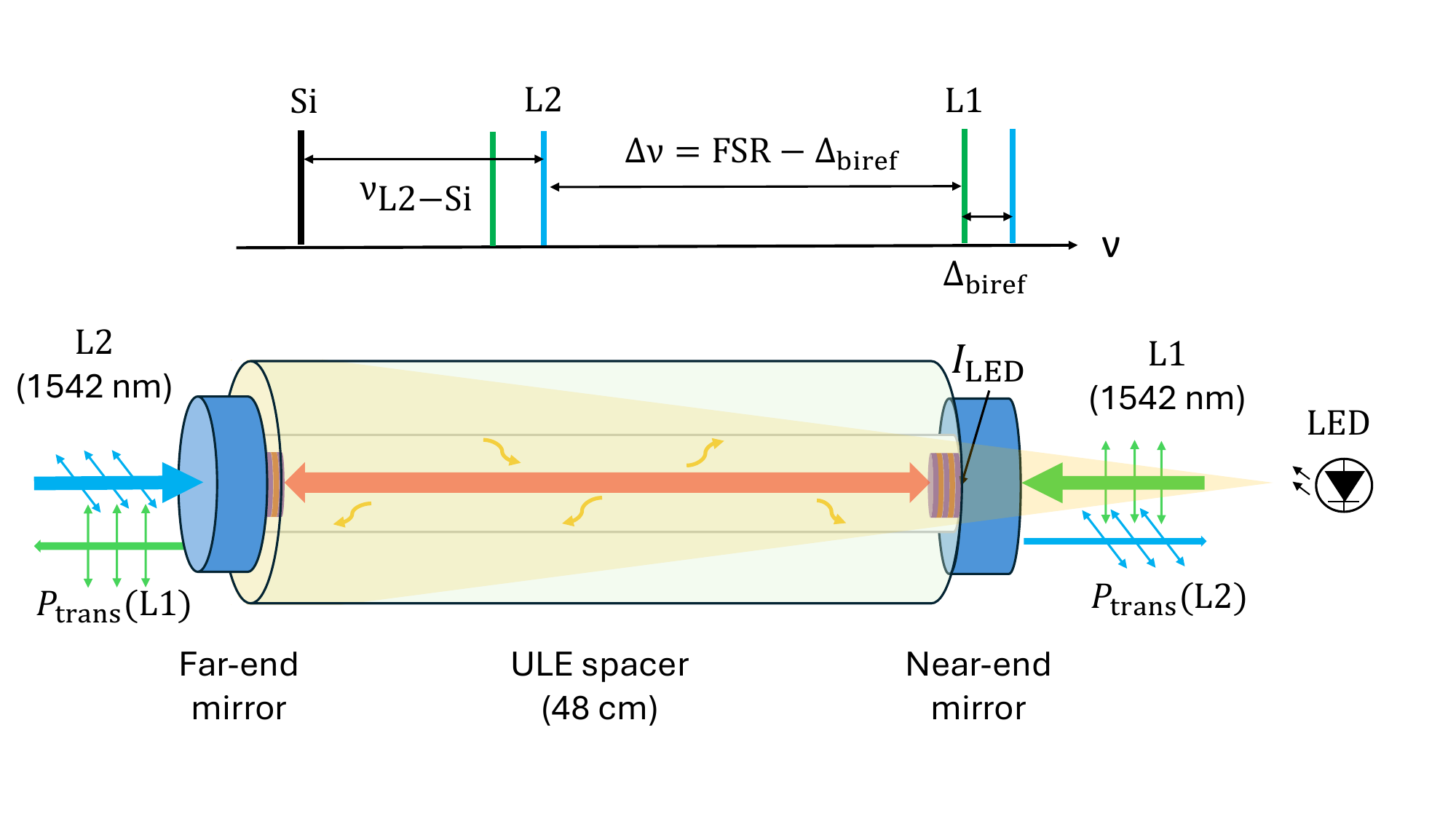}
	\caption{ 
		\label{Fig:setup}
		Experimental scheme. Two lasers (L1 and L2) at 1542\,nm are locked to eigenmodes of the ULE cavity separated by one free spectral range (FSR). 
		Their polarizations are aligned with the slow and fast axes of the mirrors respectively. 
		$P_\mathrm{trans} (\text{L1/L2})$ 
		is the transmission power from individual lasers and $I_\mathrm{LED}$ is the LED intensity at the near-end mirror. 
		The absolute frequency of L2 can be obtained from an optical beat with a laser stabilized to a Si cavity (Si).
		A LED illuminates the back side of the AlGaAs coating and enters the cavity. 
		The wavy arrows indicate scattered LED light.      
	}
\end{figure}

Fig.\,\ref{Fig:setup} shows the experimental scheme. 
Light from LEDs with wavelength of $450\pm10\,\mathrm{nm}$, $535\pm13\,\mathrm{nm}$, $625\pm7\,\mathrm{nm}$ or $890\pm22\,\mathrm{nm}$ (where the quantity after $\pm$ indicates the half width at half maximum of the emission spectrum) enters the vacuum system through one of the optical windows used for coupling laser light to the cavity. It passes two more windows on the heat shields and illuminates the rear side of the AlGaAs coating through the fused-silica substrate. 
The coating transmits 72\% of the 890-nm LED light and is opaque for shorter wavelengths.
Light can also bypass the coating and enter the cavity. This light can reach the front surface of the far-end coating directly or after scattering inside the bore hole. The front surface of the near-end mirror can also be illuminated by the scattered light.

From the geometry and the optical transmission of all three windows we can estimate the intensity $I_\mathrm{LED}$ at the back of the near-end AlGaAs coating. We choose this quantity to characterize the optical intensity in our measurements.


%

\label{sec:steady-state}

Previously we presented preliminary data of the sensitivity of birefringence to 1542\,nm intracavity light or to diffuse LED light \cite{ma24a}. We found a much higher sensitivity for light above the bandgap of GaAs.
In addition, we have observed that the birefringent splitting $\Delta_\mathrm{biref}$ from LED illumination at constant intracavity power can be described by a generic curve that only depends on the scaled intensity \cite{ma24a}.
Here we develop a comprehensive description of the birefringence modification by the simultaneous actions of 1542\,nm intracavity light and diffuse LED light. To this end we measure the birefringence response to different intracavity power, LED intensity and LED wavelength. 
The LED intensity is described by the intensity $I_\mathrm{LED}$ at the mirror that is close to the LED \cite{ma24a}.
We use the sum of transmission power from the two independent lasers L1 and L2 $P_\mathrm{trans} = P_\mathrm{trans}(\text{L1}) +
P_\mathrm{trans}(\text{L2})$ as proxy for the intracavity power $P_\mathrm{int}$ =  $P_\mathrm{trans}/T$ where $T = 15\,\mathrm{ppm}$ is the design mirror transmission.

It was observed that the behavior of birefringent modification by intracavity and by short wavelength LED illumination as function of intensity is strongly different \cite{ma24a}. We advance the hypothesis that light with photon energies below the GaAs bandgap modifies the line splitting by an initial two photon absorption process where the effect scales with the square of the optical power, in contrast to single photon absorption for above bandgap light. 
The two-photon process can be a direct or a sequential two photon absorption via intermediate states in the bandgap \cite{pen89, hur07}. 

\begin{figure}
	\includegraphics[width=\columnwidth]{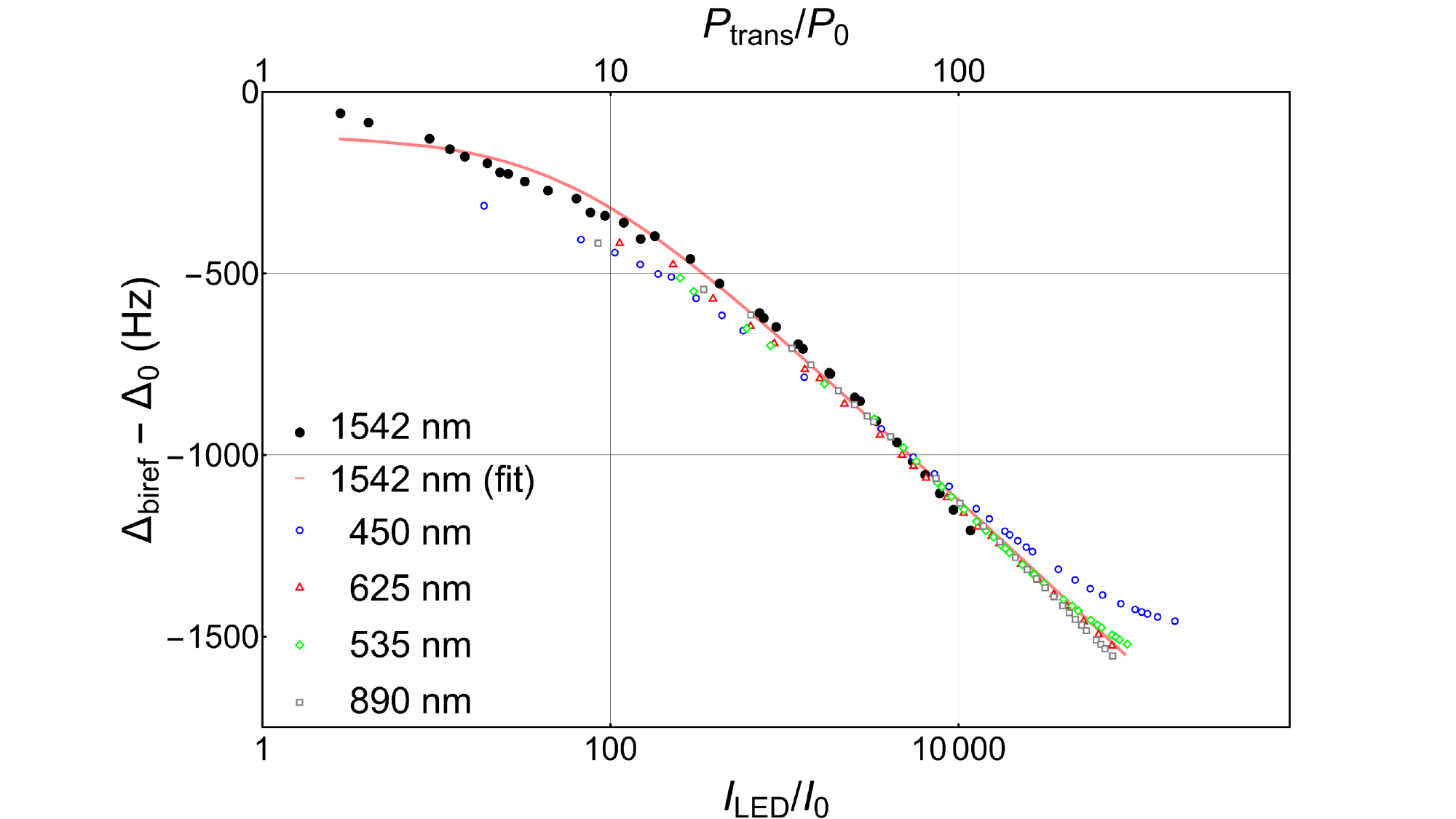}
	\caption{ 
		\label{Fig:overlap}
		Birefringent linesplitting $\Delta_\mathrm{biref}$ 
		as a function of normalized transmitted power $P_\mathrm{trans}/P_0$ (black dots, top axis) with $P_0=1\,\upmu$W, 
		and by diffuse LED light as a function of the scaled LED intensity (bottom axis) at $P_\mathrm{trans} = 7\,\upmu$W.
		Values of the bottom axis are top axis values squared.
		The red curve shows a fit of Eq. \ref{Eq:Shockley} to the 1542\,nm data.
		The scaling factors for LED amount to $I_0 = 14.3, 5.7, 3.4$ and $2.9\, \mathrm{\upmu W/m^{2}}$ at the wavelength of 
		450\,nm (blue circles),
		625\,nm (red triangles),
		535\,nm (green rhombs), and
		890\,nm (gray squares).
		For clarity, a fixed offset of $\Delta_0 = 104.280$\,kHz was subtracted in the displayed y-axis.
	}
\end{figure}

To test this hypothesis, we plot the data as function of $(P_\mathrm{trans}/P_0)^2$ and $I_\mathrm{LED}/I_0$ in the same graph (Fig.~\ref{Fig:overlap}). 
Without being able to derive the sensitivity coefficients from first principles, we can achieve a good agreement by applying the LED wavelength-dependent coefficient ${I_0}$, such that 
${I_\mathrm{LED}}={I_0}$ 
leads to a shift equal to the one induced by an intracavity power corresponding to 
$P_\mathrm{trans} = P_0 = 1\,\upmu\mathrm{W}$ 
(equivalent to intracavity intensity of $8.6 \times 10^4\,$W/m$^{2}$ on the plane mirror).


%
%

The observed correspondence between LED intensity and intracavity power squared indicates that the birefringence change can be modeled with two different initial processes driving a common mechanism.
Subsequently the birefringence could be modified via the linear electro-optic effect in GaAs \cite{ada92} from local electric fields or by the reversible change of internal strain in the coatings upon illumination \cite{ada83} (photo-elastic effect) due to recombination of charge carriers at dislocations.
The latter would induce a change of  birefringence via a modification of the stress in the coating thought to be responsible for the static birefringence \cite{win21}.

We assume, that the observed change in birefringence can be expressed by a function $\Delta_\mathrm{biref}(g)$ with a single argument proportional to the charge carrier generation rate
\begin{equation}
	\label{Eq:ModTotal}
	g = \frac{I_\mathrm{LED}}{I_0(\lambda)} + \frac{P_\mathrm{trans}^2}{P_0^2}.
\end{equation}

The logarithmic dependence of the splitting with LED intensity  (Fig.\,\ref{Fig:overlap}) over more than three decades resembles the open circuit voltage-to-power dependence of a photodiode \cite{kus17} that is described by the Shockley diode equation \cite{sho49}. 

In the coating the charge carriers (electrons in the conduction band) created in the material can diffuse to potential wells in the conduction band formed at the GaAs/AlGaAs heterojunctions between the layers \cite{fuj21, nat86} or at the surface \cite{lin17}. The charges trapped in the potential wells create electric fields that act back on the charge movement and also modify the birefringence by the electro-optic effect \cite{ada92}.  
The model presented here is similar to the one presented in ref \cite{wu25c} by considering a balance between charge generation and loss where the charge density modifies the birefringence. While their model considers spontaneous and light-induced relaxation of the charge density, we are considering an equilibrium between charge drift and diffusion. This is required to describe our observations over a large range of intensities. 
In analogy with the Shockley model, we describe the line splitting change as a function of the charge carrier generation rate  $g$: 
\begin{equation}
	\label{Eq:Shockley}
	\Delta_\mathrm{biref}(g) =
	\Delta_\mathrm{biref}^0 + 
	\Delta_s \ln \left( \frac{g}{g_s} + 1 \right),
\end{equation}
where $\Delta_\mathrm{biref}^0$ is the dark value of the line splitting. 
$\Delta_\mathrm{s}$ and $g_\mathrm{s}$ are scaling factors for the induced shift and the light related charge carrier generation.
A fit of this model to the shift $\Delta_\mathrm{biref}$ induced by 1542\,nm intracavity light is shown in Fig.\,\ref{Fig:overlap}. 
From the fit we obtain 
$\Delta_\mathrm{biref}^0 = 104.26 (1)\,\mathrm{kHz}$, 
$\Delta_\mathrm{s} = -193 (7)\,\mathrm{Hz}$, 
and $g_\mathrm{s} = 56 (10)$.

Because the LED illuminates the two mirrors quite differently,
we need to separate the contribution of the two mirrors to the modification of birefringence.
We assume that the light-induced modification of the birefringence in the mirror material is mostly a local effect, which implies a dependence on the local intensity at the mirror for both the intracavity 1542\,nm light and the LED light. 
Assuming similar sensitivity of each mirror to the local intensities, the observed splitting is the sum of the two mirror contributions:
\begin{equation}
	\label{Eq:Modfit}
	\Delta_\mathrm{biref}(P_\mathrm{trans}, I_\mathrm{LED})  = \Delta_\mathrm{biref}^0  
	+ \Delta_s \left( \ln \left( \frac{g_\mathrm{near}}{g_s} + 1 \right) 
	+ \ln \left( \frac{g_\mathrm{far}}{g_s} + 1 \right) \right)
\end{equation}

with
\begin{equation}
	\label{Eq:Modfit2}
	\begin{split}
		g_\mathrm{near} &= \hphantom{\rho}   P_\mathrm{trans}^2/P_0^2    + \hphantom{\eta}  I_\mathrm{LED}/I_0(\lambda) \\
		g_\mathrm{far}  &=     \rho          P_\mathrm{trans}^2/P_0^2 +        \eta         I_\mathrm{LED}/I_0(\lambda).
	\end{split}
\end{equation}

For the intracavity light the $1/e^2$ mode radii $w$ at the two mirrors differ significantly ($w_\mathrm{concave} = 687\,\mathrm{\upmu m}$, $w_\mathrm{plane} = 496\,  \mathrm{\upmu m}$) leading to ratio of the intensities of 1.92 and thus to a factor $\rho=1/3.7$ in the generation rate $g$.
The influence of LED light on the far-end mirror is described by scaling the contribution by a factor $\eta$ compared to the one at the near end mirror.

\begin{figure}
	\includegraphics[width=\columnwidth]{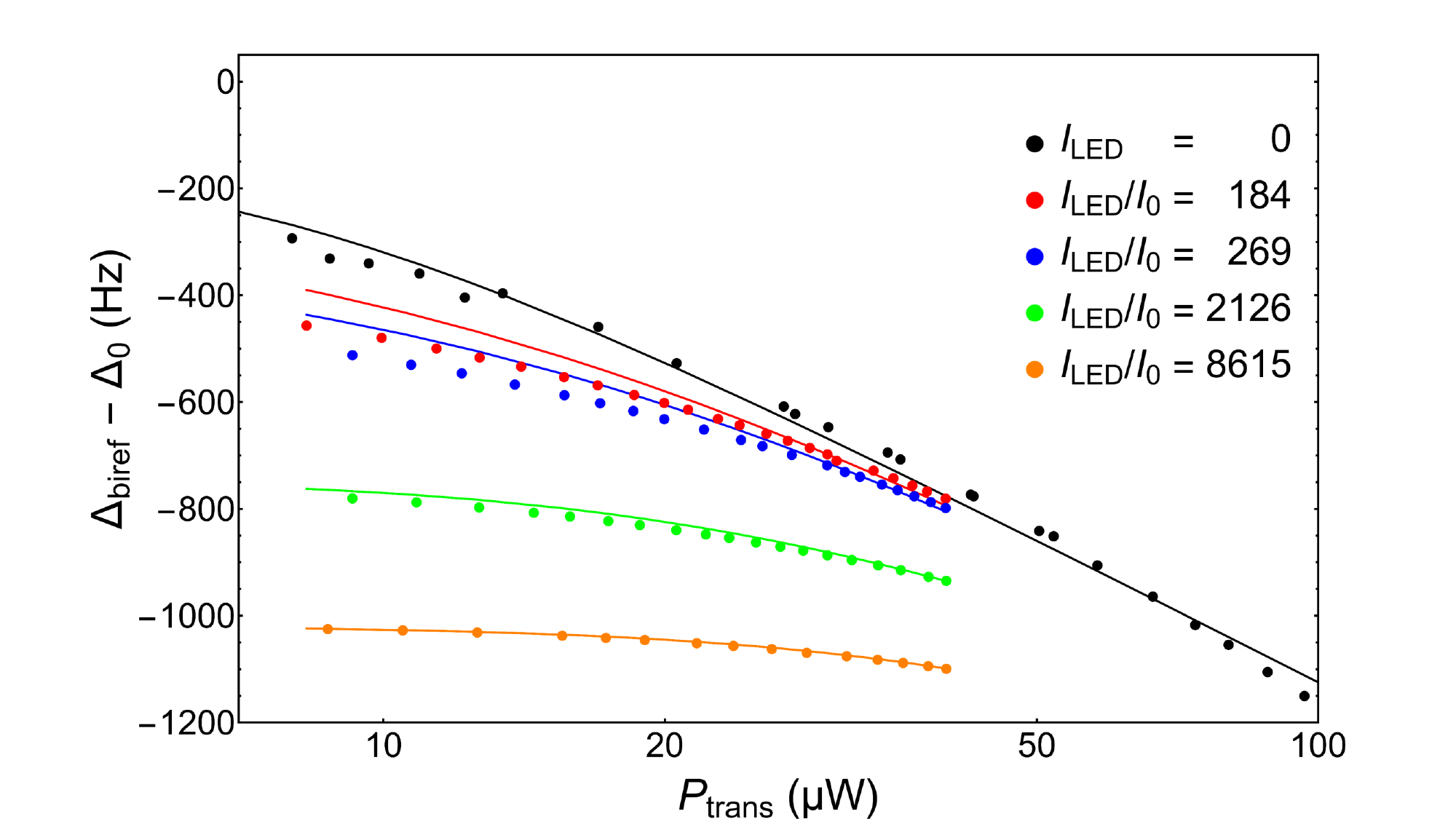}
	\caption{
		\label{Fig:Staticsum}
		Birefringent line spitting $\Delta_\mathrm{biref}$ as a function of transmission power $P_\mathrm{trans}$ at different intensities of a green LED emitting at 535\,nm. 
		Eq.\,\ref{Eq:Modfit} was fitted to data at the highest LED intensity $I_\mathrm{LED} / I_0$ = 8615 (orange line) to determine the ratio of effective LED intensities at both mirrors.
		The curves (green, blue and red) show the model using this ratio. 
		For clarity, a fixed offset of $\Delta_0$ = 104.280 kHz was subtracted in the displayed y-axis.}
\end{figure}

We measure $\Delta_\mathrm{biref}$ as a function of intracavity power at various LED intensities (Fig.\,\ref{Fig:Staticsum}). 
The rear side of the plane mirror was irradiated with a green LED $(\lambda = 535\,\mathrm{nm})$, where $I_0(535\,\mathrm{nm}) = 3.4 \,\mathrm{\upmu W/m^2}$ was determined previously. 

First, we fit the model (Eqs \ref{Eq:Modfit}, \ref{Eq:Modfit2}) to the observed shift at the highest LED intensity (Fig. \ref{Fig:Staticsum}, orange circles).
From the fit, the sensitivity to LED light 
of the far-end mirror is a factor of $\eta = 5.1 (1)$ bigger than the one of the near-end mirror. 
Taking the different intensities into account, this indicates that the coatings are much more sensitive to front-side illumination.

With this ratio, we model the response to $P_\mathrm{trans}$ at other LED intensities using Eq. \ref{Eq:Modfit}, as shown in Fig. \ref{Fig:Staticsum}. 
The fair agreement between the experimental data and our simple model supports the assumption that effects from both intracavity light and external LED light simply add in the charge carrier generation rate $g$ that non linearly modifies the line splitting. 

\section{Transient response to intracavity power}
\label{sec:transient}
The temporal response of $\Delta_\mathrm{biref}$ to a step change in intracavity power provides further insight into the mechanisms.
Previously, it was observed that the transient response of $\Delta_\mathrm{biref}$ strongly varies with temperature, and it is faster at higher final intracavity power \cite{yu23a, kra23a, zhu23,ma24a}. 
Here, we further investigate the temporal photo-birefringent response at 1542\,nm with or without constant LED light in the visible and near infrared. 

As the magnitude of the change depends on the step size, we normalize all the curves by the difference of initial $\Delta_\mathrm{biref}(t_0)$ and final line splitting $\Delta_\mathrm{biref}(t_\mathrm{final})$

\begin{equation}
	\label{Eq:delta_biref}
	\delta_\mathrm{biref}(t) = \frac{\Delta_\mathrm{biref}(t) - \Delta_\mathrm{biref}(t_\mathrm{final})}
	{\Delta_\mathrm{biref}(t_0) - \Delta_\mathrm{biref}(t_\mathrm{final})}, 
\end{equation}
resulting in a normalized response of birefringence from 1 to 0 that was found to be independent of the initial power level and the step size of the power \cite{ma24a}.

\subsection{Without additional LED illumination}
The transient responses of $\delta_\mathrm{biref}$ to step changes of intracavity power is investigated. 
The half time of the transient response decreases from 1.3\,s to 0.3\,s when the final power $P_\mathrm{trans}$ is increased from 10\,$\upmu$W to 49\,$\upmu$W. 
Different from the observed temporal behavior at 124\,K \cite{yu23a}, here the splitting is monotonically approaching its final value. 
We find that the transient behavior can be best described by a sum of an exponential term representing the initial relaxation and a stretched exponential term that describes the slow part of the curve Eq.\,\ref{Eq:transfit}:
\begin{equation}
	\label{Eq:transfit}
	\delta_\mathrm{biref}(t) =
	A\, e^{-(\alpha t/\tau_1)^{\beta}} + (1-A)\, e^{-\alpha t/\tau_2}.
\end{equation}
First, a fit to the data at $P_\mathrm{trans}= 10\,\upmu$W ($\alpha=1$) was performed,  with weighting factor of $1/\delta_\mathrm{biref}(t)$ to ensure a good fit across the whole decay. 
The fit gives a best exponent $\beta$ = 0.510(1) and relative amplitude $A = 0.579(4)$. 
The time constants of the stretched exponential and the normal exponential are $\tau_1$ = 2.92(6)\,s and $\tau_2$ = 1.65(1)\,s, respectively. 
The stretched exponential behavior of $\delta_\mathrm{biref}$ might be related to the observations of slow relaxation of photo carriers in persistent photo-conductivity of AlGaAs, 
where the exponent $\beta$ was found to be temperature dependent \cite{gho04}.   

Next, only $\alpha$ in Eq.\,\ref{Eq:transfit} was fit to the measurement at other  $P_\mathrm{trans}$ with the other parameters fixed to the result of the initial fit. This corresponds to stretching the time axis by a factor of $\alpha$ to overlap all the curves  (see  Fig.\,S1 in the Supplementary Material).  
Fig.\,\ref{Fig:Transcavscale} shows two curves ($P_\mathrm{trans}$ = 49 $\upmu$W and 28 $\upmu$W) with scaled time axis $\alpha \,t$ and the original curve at 10\,$\upmu$W that well overlap. 
The scaling factors $\alpha$ for a total of ten different $P_\mathrm{trans}$ with respect to the curve at final $P_\mathrm{trans}= 10\,\upmu$W is shown in the inset of Fig.\,\ref{Fig:Transcavscale}. The uncertainties were obtained from fits with different weighting factors (see Supplementary Material). A linear fit to the scaling factor results in $\alpha  = 0.102(1) \cdot P_\mathrm{trans}/\upmu\mathrm{W} - 0.03(3)$. 

%
\begin{figure}
	\includegraphics[width=\columnwidth]{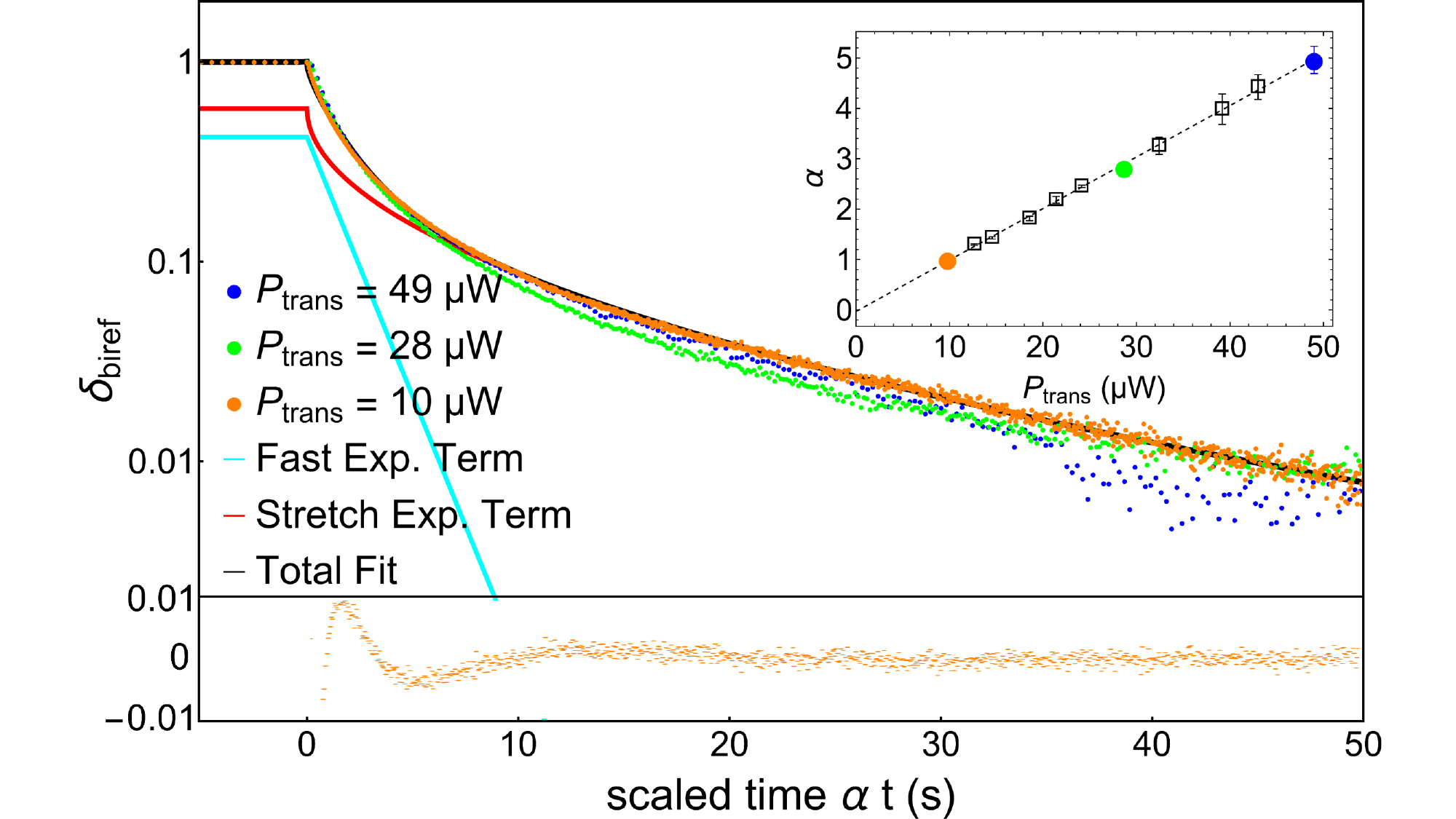}
	\caption{	\label{Fig:Transcavscale}
		Normalized transient response $\delta_\mathrm{biref}$ to a step change of intracavity power without LED illumination. Responses of three final transmitted powers of 10\,$\upmu$W, 28\,$\upmu$W and 49\,$\upmu$W are shown. 
		The time axes of the two curves at $P_\mathrm{trans}$ = 28 and 49 $\upmu$W are scaled by $\alpha = 2.8$ and $\alpha = 5.0$ with respect to the slower curve at 10\,$\upmu$W (orange) ($\alpha = 1$). 
		The response at 10\,$\upmu$W was fitted according to Eq.\,\ref{Eq:transfit} with a sum of an exponential function (cyan) and a stretched exponential function (red)
		representing the slow part. 
		The fit residuals are displayed in the lower panel.
		The inset displays the scaling factors $\alpha$ versus the final power $P_\mathrm{trans}$ for the curves in the main figure (circles) and at other power levels (open squares).
	}
\end{figure}
\subsection{With additional constant LED illumination}
\begin{figure}
	\includegraphics[width=\columnwidth]{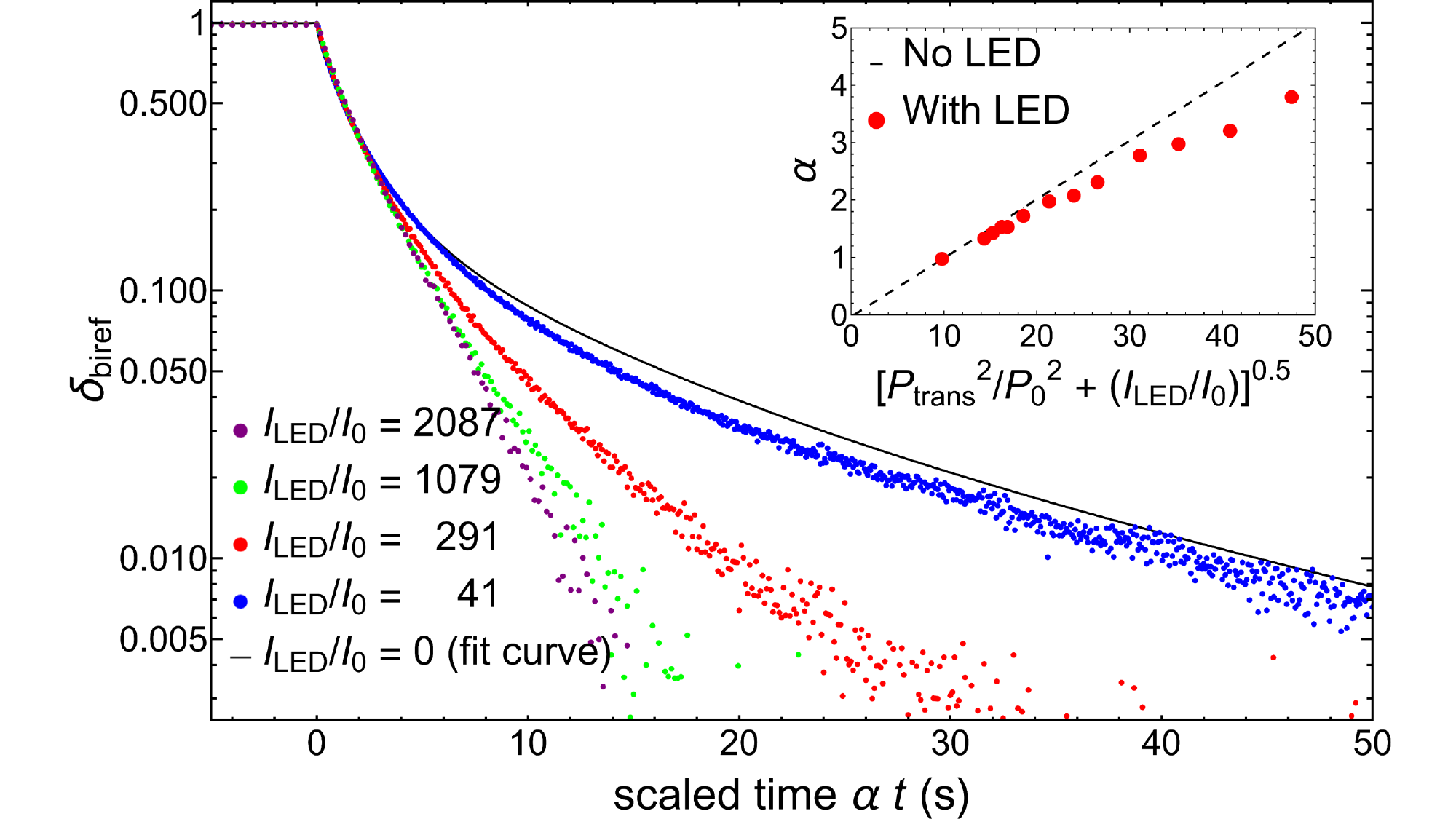}
	\caption{\label{Fig:TranscavLEDscale}
		Normalized transient response $\delta_\mathrm{biref}$ to a step change of intracavity power (corresponding to $P_\mathrm{trans} = 11.0\, \upmu$W to 12.8 $\upmu$W) at different values of $I_\mathrm{LED}$ at 535\,nm.
		The time axis is scaled by a factor $\alpha$ compared to the axis without LED illumination (see text). The same step change in intracavity power without LED is shown as reference (black).
		The inset displays $\alpha$ as a function of transmitted power $P_\mathrm{trans}$ and LED intensity $I_\mathrm{LED}$ at fixed $P_\mathrm{trans} = 12.8\,\upmu$W (red) and without LED, varying $P_\mathrm{trans}$ (black dashed as in the inset of Fig.\,\ref{Fig:Transcavscale}). 
	}
\end{figure}

Next we investigate the temporal response to a change in intracavity power, when the mirrors are additionally illuminated by LED light at fixed intensities. 
The same 535\,nm green LED is used as before.
We measure the transient responses of the line splitting to a change of intracavity power, corresponding to $P_\mathrm{trans}$ = 11.0\,$\upmu$W to 12.8\,$\upmu$W. 
We try to overlap the normalized responses obtained at different LED intensities (see Fig.\,S2 in the Supplementary Material) by stretching the time axis. We observe that the curves cannot be overlapped: the long term behavior is very different. Nevertheless,
we can overlap the curves at short times by stretching the time axis such that they agree at $\delta_\mathrm{biref} (\alpha\; t) = 1/e $ (see Fig.\,\ref{Fig:TranscavLEDscale}), where  $\alpha = 1 $ refers to the same reference curve without LED as in Fig.\,\ref{Fig:Transcavscale}.

This common reference allows us to compare the dependence of the scaling factors $\alpha$ on intracavity power and LED intensity (inset of Fig.\,\ref{Fig:TranscavLEDscale}). 
The factors are plotted versus the square root of quantity $g$ (Eq.~\ref{Eq:ModTotal}) that combines also the effects of intracavity power and the LED intensity in the steady state modification of the splitting. 
Both sets of $\alpha$ agree well at small values of $\sqrt{g}$. 
The deviation at high $I_\mathrm{LED}$ may be related to the different behavior of the two mirrors due to the different intracavity light and LED light intensities impinging on each.

\subsection{Transient model}
We try to extend the steady state model to describe the transient response. 
Similar to reference \cite{wu25c}, we describe the observed splitting from the charge $\rho$ under constant illumination as an equilibrium between charge generation rate $G$ and loss $L$, assuming a spatially homogeneous behavior:
\begin{equation}
	\frac{d}{dt}\Delta_\mathrm{biref} \propto \ \frac{d \rho}{dt} = G - L.
\end{equation}
Linearizing this equation for small deviations $\delta\rho$ from the equilibrium value at $G=L$, we arrive at
\begin{equation}
	\frac{d}{dt} \delta\rho = - \frac{\partial L}{\partial \rho} \cdot \delta\rho .
\end{equation}
The solution of this differential equation is an exponential function with time constant $\Gamma = \frac{\partial L}{\partial \rho} $.
At equilibrium 
\begin{equation}
	\Gamma = \frac{\partial L}{\partial \rho} = \frac{\partial G}{\partial \rho} = \left(\frac{\partial \rho}{\partial G}\right)^{-1} .
\end{equation}
For the Shockley model of Eq. \ref{Eq:Shockley}, $\rho \propto \ln(g/g_s + 1)$. At equilibrium, using the generation rate derived from the steady state results presented in Sec. \ref{sec:steady-state} $G \propto g = I_\mathrm{LED}/I_0 + (P_\mathrm{trans}/P_0)^2$, we obtain $\Gamma \propto I_\mathrm{LED}/I_0 + (P_\mathrm{trans}/P_0)^2$ and
\begin{equation}
	L(\rho) \propto e^{\rho/\rho_0}-1.
\end{equation}

This model does not match the observed acceleration of the transient response which is better described by $\Gamma \propto \sqrt{I_\mathrm{LED}/I_0 + (P_\mathrm{trans}/P_0)^2}$.
We have no clear explanation, one may speculate that the spatial inhomogenities from the radial Gaussian profile and along the depth of the coating play important roles. We have also not considered the dynamics of charge separation leading to the buildup of space charges. 

\section{Transient response to change of LED intensity}

\begin{figure}
	\includegraphics[width=\columnwidth]{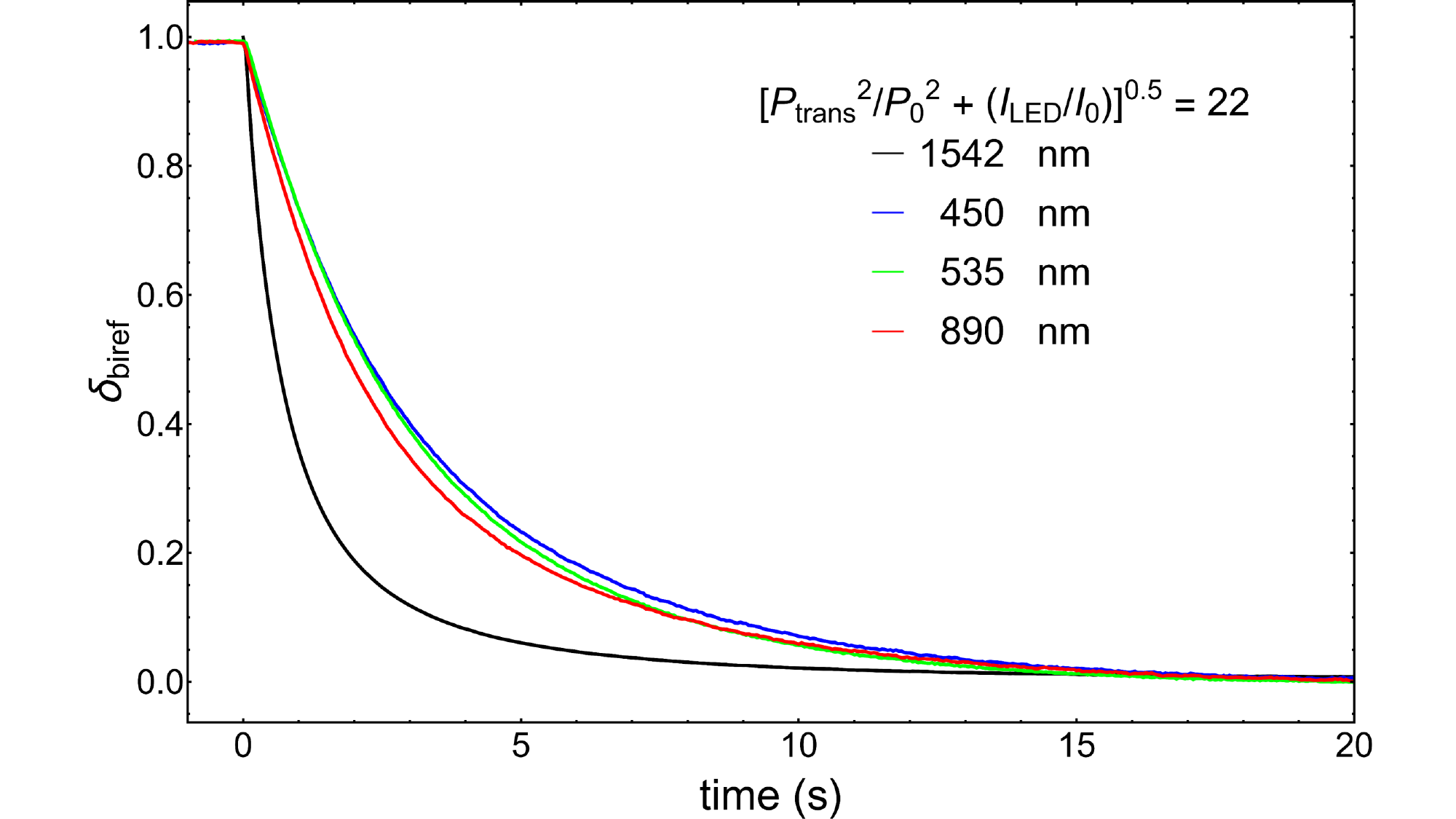}
	\caption{\label{Fig:Transinandout}
		Normalized transient response of $\delta_\mathrm{biref}$ to switching on different LEDs at 450\,nm (blue), 535\,nm (green) and 890\,nm (red). 
		All measurements have the same intracavity power corresponding to $P_\mathrm{trans}=  10.7\,\upmu$W and the LED intensity was adjusted to lead to the same final splitting $\Delta_\mathrm{biref}$.	
		For comparison the response to a step of intracavity power is shown by the black line.
		Here the power was stepped up from $P_\mathrm{trans}= 10.7\,\upmu$W to $22.0\,\upmu$W, leading to the same final splitting. 
	}
\end{figure}

Finally, the temporal response of the photo-modified birefringence to a step of LED intensity at constant intracavity power was studied. 
Here also higher LED intensity results in a faster response (see Fig.\,S3 and S4 in Supplementary Material), similar to the behavior on intracavity power.
We measure the responses of the normalized birefringent splitting to turning on LEDs (IR, green and blue) at constant intracavity power $P_\mathrm{trans}$ = 10.7\,$\upmu$W and compare them to a step of intracavity power from $P_\mathrm{trans}$ = 10.7\,$\upmu$W to 22.0\,$\upmu$W without LED (Fig.\,\ref{Fig:Transinandout}). 
The LED intensity was chosen to result in the same final splitting, independent of the wavelength.
According to our model this corresponds to setting $\sqrt{g}$ = 22 (Eq.~\ref{Eq:ModTotal}). 
The observed responses to LED light are much slower than the response to intracavity light (half time 0.7\,s).
This could be related to radial charge transport from the whole illuminated coating surfaces with radius of 4\,mm to the area of the cavity mode at the mirrors with beam radii of 687\,$\upmu$m and 496\,$\upmu$m.

The small difference of the temporal response of $\Delta_\mathrm{biref}$ to a step change in intracavity power between the near band gap light (halftime of 2.0\,s) and blue/green light (halftime of 2.3\,s) could be due to the different depths where the charges are created.  
The absorption spectrum of GaAs \cite{asp86a} leads to predominant absorption of the blue and green LED light in the first few layers, whereas 890\,nm LED light is absorbed throughout all coating layers (see Supplementary Table S1). 

\section{Reduction of sensitivity to laser power fluctuations}

\begin{figure}
	
	\includegraphics[width=\columnwidth]{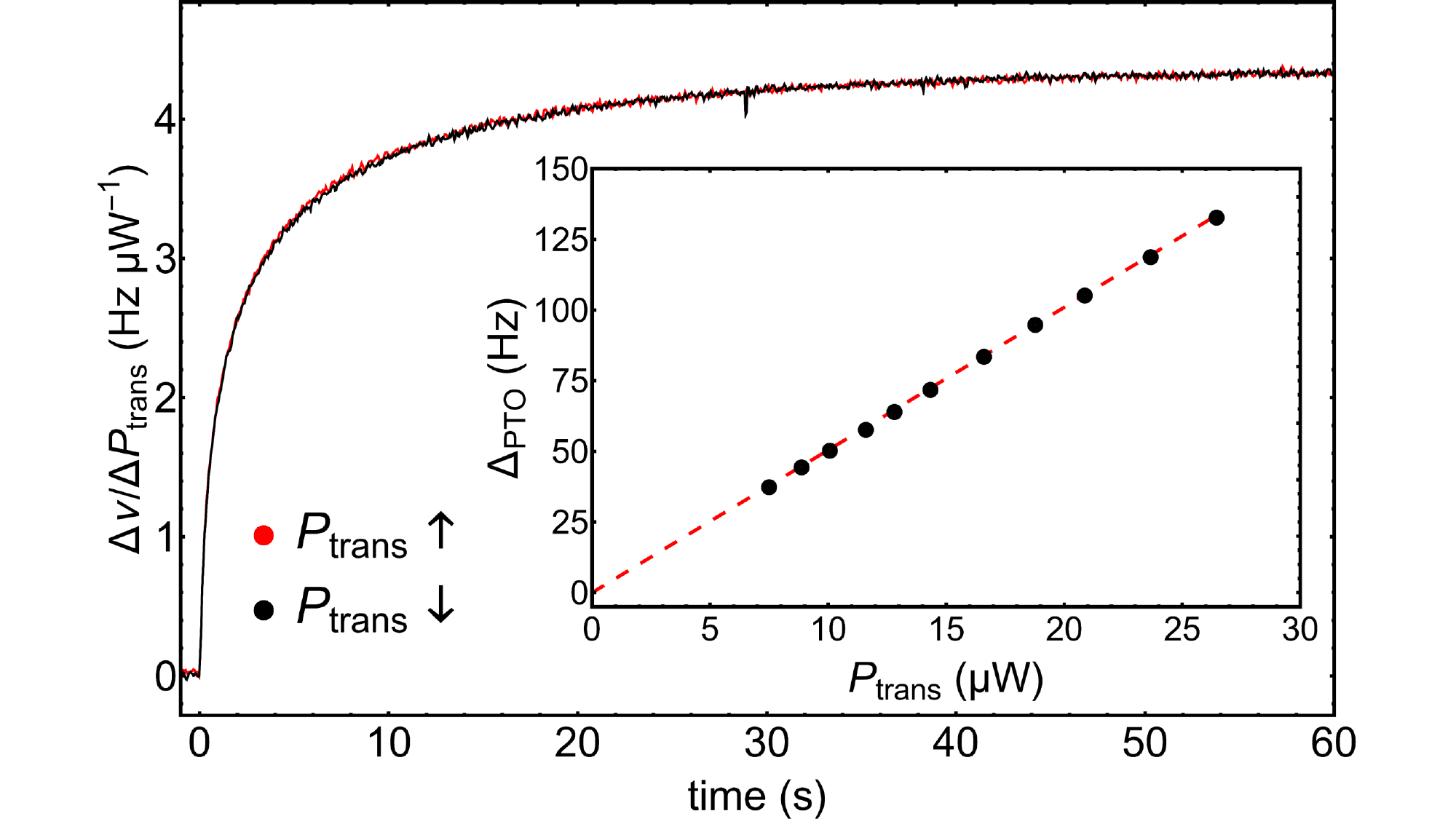}
	\caption{\label{Fig:Commonmode}
		Transient change of the average frequency of a fast and a slow cavity eigenmode $\Delta_\mathrm{PTO}$ to a step of intracavity power from $P_\mathrm{trans}=9.2\,\upmu$W  up to $10.9\,\upmu$W (red) and back down to $9.2\,\upmu$W (black), normalized by the step size.  
		The inset shows the steady-state frequency changes $\Delta_\mathrm{PTO}$ as function of $P_\mathrm{trans}$ extrapolated to 0 with a linear fit shown in the dashed line. 
	}
\end{figure}

Unavoidable absorption of the intracavity light leads to heating of mirror coatings and the mirror substrates underneath. 
The thermal expansion of the coating and the substrate (thermo-elastic effect) and the temperature dependent index of the coating material (thermo-refractive effect) will shift the resonance frequency of the cavity by $\Delta_\mathrm{PTO}$ (subsumed as photo-thermo-optic (PTO) effect) \cite{far12}. 
Thus fluctuations of intracavity power lead to frequency fluctuations.
To measure the PTO effect, we observe the average frequency change of the fast and the slow eigenmodes to a step in optical power (see Fig.\,\ref{Fig:Commonmode}). 
Because of the thermal expansion of the fused silica substrates, the cavity length shortens with increasing power.
As expected from a pure thermal effect, the amplitude of the transient response from $\Delta_\mathrm{PTO}$ is proportional to the step size of optical power (Fig.\,\ref{Fig:Commonmode} inset) with slope 4.5\,Hz/$\upmu$W, equivalent to an optical length change of 5.6\,fm/$\upmu$W for each mirror.
In addition, the shape of the transient response, normalized to step size is independent of power, with 1.3\,s halftime, similar to other cavities with AlGaAs coatings on fused silica substrates \cite{her21_short,zhu23}.

Fluctuations of the intracavity power couple to the cavity frequency $\nu$ via the PTO effect \cite{ros02} and via the modification of the birefringence simultaneously. 
As the birefringence shows opposite signs on fast and slow axes, the corresponding frequency on the fast and slow axes is given by 
\begin{equation}
	\label{Eq:nu_sum}
	\nu_\mathrm{fast,slow} = \nu_0 + \Delta_\mathrm{PTO} ~ \pm ~ \Delta_\mathrm{biref}/2,
\end{equation}
where $\nu_0$ is the unperturbed frequency.
Because the birefringent splitting response for small changes to intracavity power depends on the absolute power, intracavity power can be chosen to obtain cancellation of the PTO and birefringent small signal response. For our system we obtain such cancellation at an transmission power of $P_\mathrm{trans} = 52\,\upmu\mathrm{W}$.
Because the time response of the two effects is generally different  the cancellation cannot be perfect.
For our system we achieve a residual power sensitivity  smaller than 
$\pm 0.5\,\mathrm{Hz}/\upmu$W for Fourier frequency larger than 0.1 Hz.

\begin{figure}
	
	\includegraphics[width=\columnwidth]{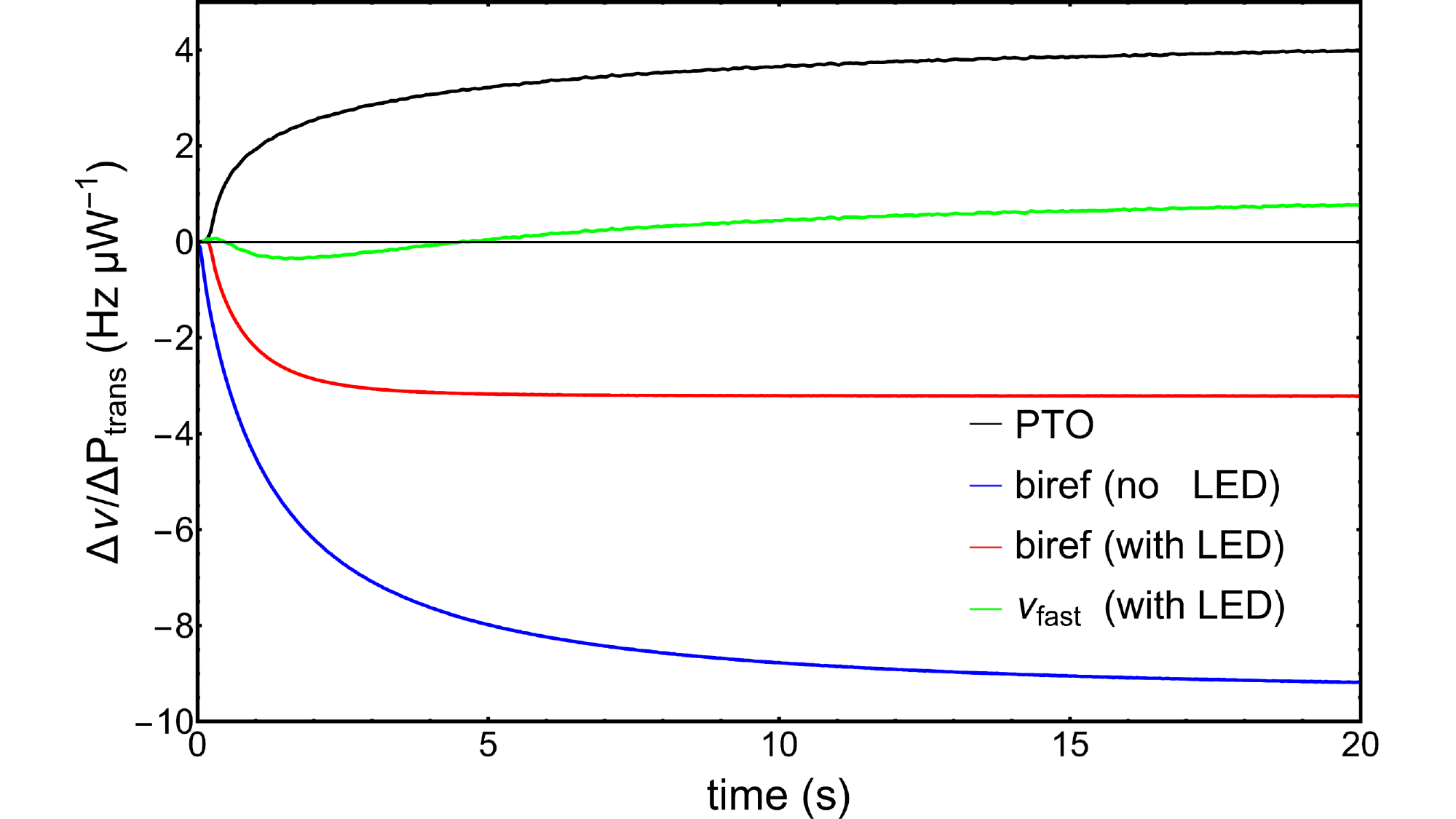}
	\caption{\label{Fig:Noiseoptimum}
		Contributions to the resonance frequency changes upon a step in intracavity power from photo-thermo-optic effect (black) and from the photo birefringent effect without (blue) and with constant LED illumination at 535\,nm (red). 
		The total shift for the fast axis with LED illumination is shown in green.
	}
\end{figure}

For fundamental (shot noise) and technical reasons  (beam pointing stability) the absolute power stability  decreases at smaller optical power. For achieving better frequency stability it is therefore advantageous to operate at a lower power.
%
%
LED illumination offers the possibility to tune the sensitivity of the photo birefringent effect to optimize the cancellation at a lower intracavity power.
Fig. \ref{Fig:Noiseoptimum} shows an example where at an intracavity power corresponding to  $P_\mathrm{trans}=11\,\upmu$W, LED illumination is used to reduce the sensitivity to power from 4\,Hz/$\upmu$W without LED light to below 0.5\,Hz/$\upmu$W for Fourier frequency up to 0.1\,Hz.
LED illumination is much less sensitive to  technical fluctuations like beam pointing thus it will not significantly degrade the laser stability.  

%
%

%

%
\section{Conclusion and Outlook}

From the investigations of the photo-birefringent effects of crystalline AlGaAs coatings at room temperature we have derived a model 
that is based on two processes.
In the first process light creates charges by photoabsorption. 
The modification of birefringence is much stronger for LED light above the bandgap of GaAs and also shows different dependence on power compared to intracavity light below the bandgap of GaAs. 
This suggests an initial single photon absorption by GaAs for short wavelengths, in contrast to two photon absorption for long wavelengths.

In the next process the charge carriers generated by photon absorption diffuse and modify the birefringence by the electro-optic effect or by the photo-elastic effect. 
The normalized transient response mostly depends on the final modification of the birefringence, with or without LED illumination.
This suggests that the response is limited by the dynamics of the second process, like charge diffusion, while the first process happens much faster.

Crystalline coatings offer the possibility for next generation room temperature cavities aiming at $1 \times 10^{-17}$ fractional frequency instability due to the low thermal noise floor. 
At this level, coupling of laser power noise to frequency noise is still a technical challenge.
For example, in state of the art room temperature cavities with AlGaAs coatings, laser power noise contributes to the instability up to 4~$\times 10^{-17}$ even with active power stabilization and choosing the polarization with the smaller sensitivity, where photo-thermo-optic effect and photo birefringent effect partially cancel \cite{zhu24, kra25a}. 
The better understanding of the light induced birefringence obtained in our investigations offers the possibility to achieve good cancellation at much lower laser power levels by adding LED illumination.
As the fractional power stability typically remains the same, operating at lower power thus opens a pathway towards low thermal noise limited instability. 

Extending our investigations of the photo-birefringent effects to other temperatures and with well controlled area of illumination would be useful.  
In addition, investigations on the inherent novel coating noise processes would be needed to understand the best performance of AlGaAs coatings at room temperature.

\section{Acknowledgements}
We acknowledge support by the Project 20FUN08 NEXTLASERS, which has received funding from the EMPIR programme cofinanced by the Participating States and from the European Union’s Horizon 2020 Research and Innovation Programme, and by the Deutsche Forschungsgemeinschaft (DFG, German Research Foundation) under Germany’s Excellence Strategy–EX-2123 QuantumFrontiers (Project No. 390837967), SFB 1227 DQ-mat (Project No. 274200144). 
This work is partially supported by the Max Planck-RIKEN-PTB Center for Time, Constants and Fundamental Symmetries. 
We thank Jun Ye and Dhruv Kedar for the insightful discussions.

\newpage

\section{Supplementary Material}

	The following provides additional information on the wavelength dependent absorption of the crystalline coatings and transient frequency changes due to step changes in intracavity light at 1542\,nm and to step changes of diffuse LED light ranging from 450\,nm to 890\,nm.  

	\maketitle
\subsection{Wavelength-dependent coating absorption}

We calculate the wavelength and spatially-dependent absorption in an Al$_{0.92}$Ga$_{0.08}$As/GaAs coating from absorption coefficients of GaAs and Al$_{0.8}$Ga$_{0.2}$As \cite{asp86a}. 
In the wavelength range of interest, the absorption of Al$_{0.92}$Ga$_{0.08}$As can be neglected compared to the one of GaAs.
We first calculate the absorption depth of GaAs (Table. \ref{tab:setups}). 
The coatings consist of alternating GaAs (115.6 nm) and AlGaAs layers (133.2 nm), starting with GaAs.
The blue and green light is absorbed mainly in the first two GaAs layers.
For the IR light the absorption depth is larger than the coating thickness and light is absorbed throughout the whole coating (consisting of 38.5 pairs). 74\% of the light is transmitted by the coating.

\begin{center}
	\begin{table}[h]
		\centering
		\caption{\label{tab:setups}
			Wavelength-dependent absorption in the Al$_{0.92}$Ga$_{0.08}$As/GaAs coating.
			The $1/e$ absorption depth of GaAs is shown in the second column \cite{asp86a} and the corresponding thickness in terms of layer pairs is estimated.
		}
		
		\begin{tabular}{c c c c}
			Wavelength ~  &  ~ Absorption depth ~ & ~ Coating layers\\
			\hline
			459 nm  & 0.04 $\upmu$m  & 1 pair\\
			539 nm  & 0.13 $\upmu$m  & 2 pairs\\
			885 nm  & 15.38 $\upmu$m & 62 pairs\\ 
			
		\end{tabular}
	\end{table}
\end{center}

\subsection{Transient response to intracavity power without additional LED illumination}

\begin{figure}
	
	\includegraphics[width=\columnwidth]{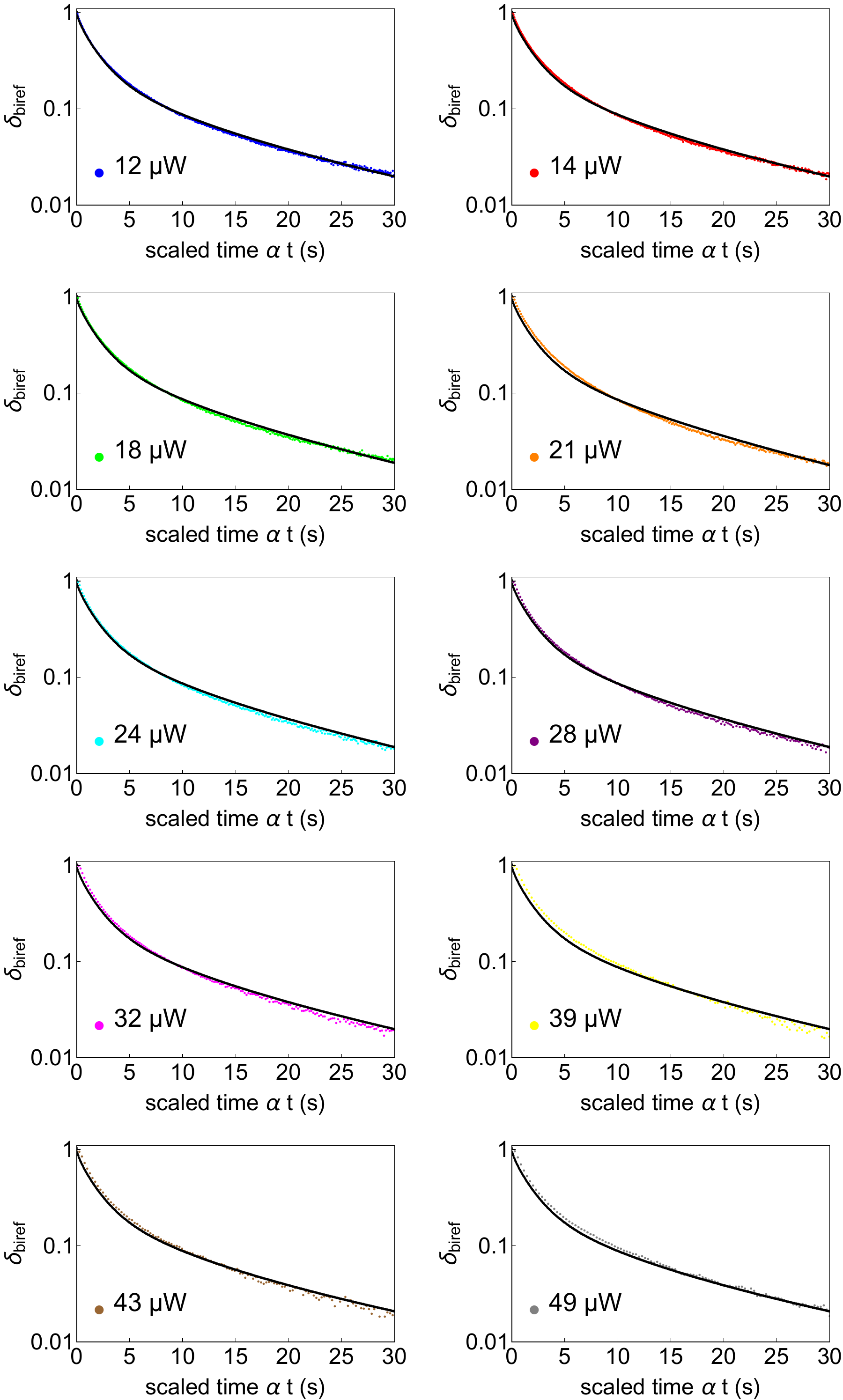}
	
	\caption{\label{Fig:Alphaverification}
		Measured normalized transient behavior of line splitting  $\delta_\mathrm{biref}(t)$ at 10 different final transmission power. The lines display the fitting results with varying (black) and constant (gray) weighting factors.
	}
\end{figure}

We use Eq. 6 to fit the stretching factor $\alpha$ to the normalized transient line splitting due to a step change in intracavity power. Especially for the faster responses at higher power we observe that we cannot equally well fit the response at short and long time. By using a constant weight or a weight of $1/\delta_\mathrm{biref}(t)$ in the fit we can achieve better agreement at short or longer times respectively. In the second case the maximum weight is limited to 100 to not give too much weight to noise. Fig. \ref{Fig:Alphaverification} displays the original $\delta_\mathrm{biref}(t)$ data and the fitted curves for the two cases. 

The inset in Fig.\,4 displays the average value for the fitted alpha in the two cases and the error bars show the deviation between the two. 

\subsection{Transient response to intracavity power with additional LED illumination} 
\begin{figure}[tbh]
	\includegraphics[width=\columnwidth]{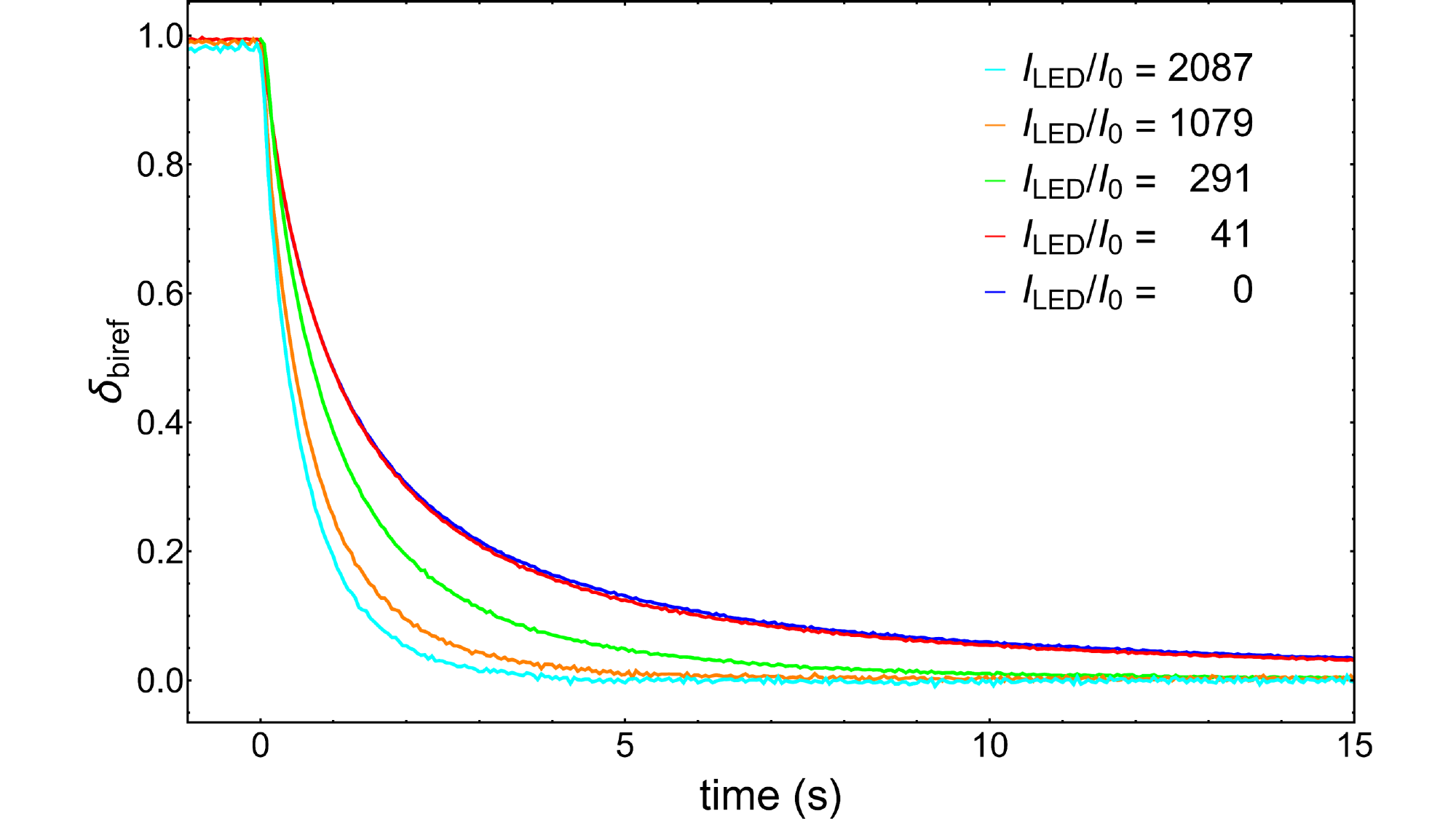}
	\caption{
		\label{Fig:TranscavLED}
		Normalized transient response $\delta_\mathrm{biref}$ to the same step of intracavity power from at 11.0 $\upmu$W to 12.8 $\upmu$W at four intensities $I_\mathrm{LED}$ of the green LED.  
		At higher intensity, the transient response is faster. The transient response without LED light (blue) is shown for comparison.
	}
\end{figure}

Fig. \ref{Fig:TranscavLED} shows the normalized transient response of a step change in cavity power ($P_\mathrm{trans}$ = 11.0\,$\upmu$W to 12.8\,$\mathrm{\upmu}$W) at different LED intensities. At a higher constant LED intensity the response is faster, similar to the case of varying intracavity power. At very small LED intensity the transient behavior is similar to the response without LED.

\subsection{Transient response to change of LED intensity}
\begin{figure}
	
	\includegraphics[width=\columnwidth]{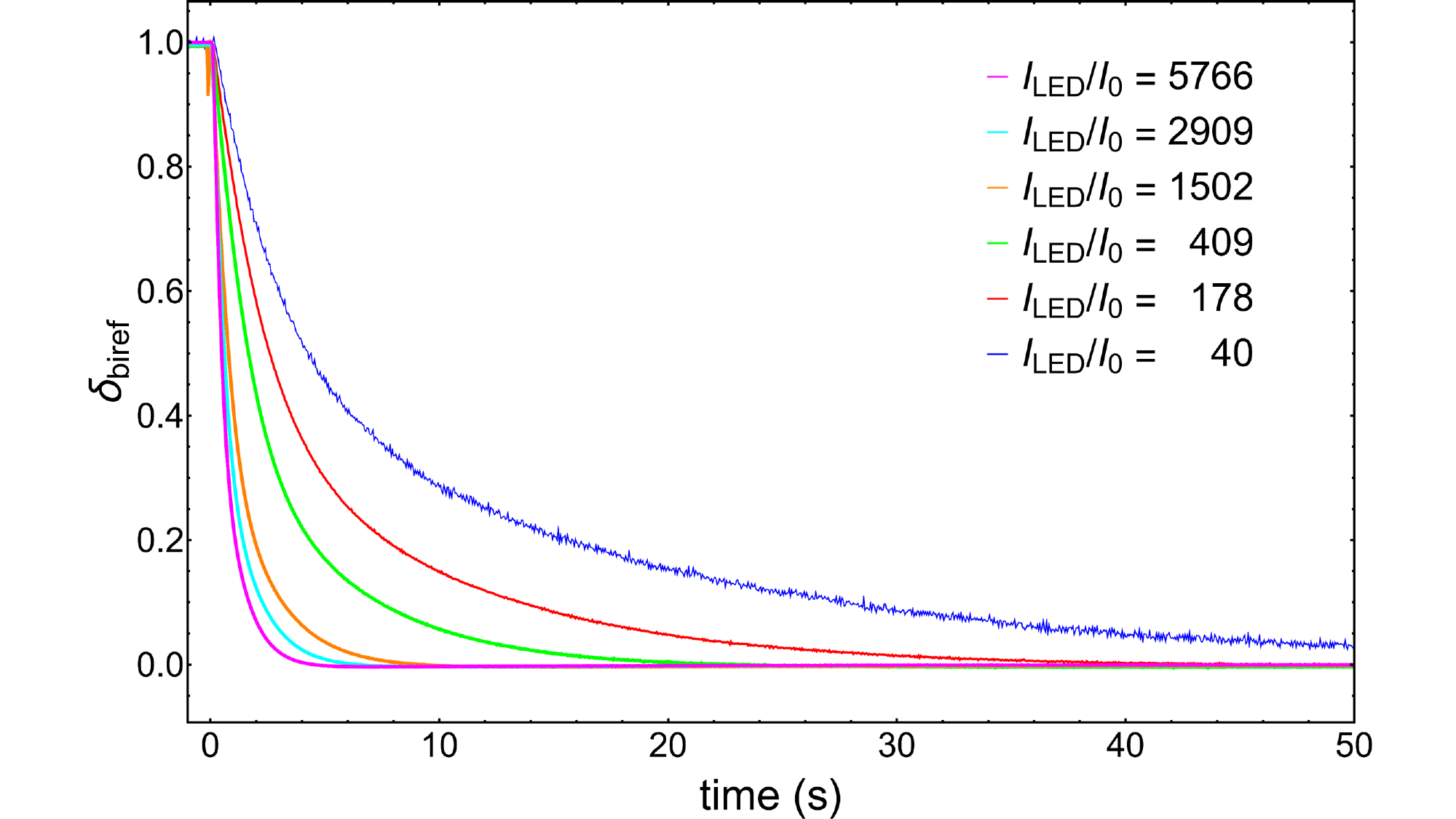}
	\caption{\label{Fig:TransLED}
		Normalized transient responses of $\delta_\mathrm{biref}$ to switching on a 535 nm  LED to different final intensities $I_\mathrm{LED}$ with $I_0 = 3.4~\upmu\mathrm{W/m^2}$.
		The intracavity power was fixed to $P_\mathrm{trans}$ = 12.5\,$\upmu$W. 
	}
\end{figure}

We also investigate the transient response to switching on a green LED for different final intensities.
The normalized transient responses are shown in Fig. \ref{Fig:TransLED}. 
Similar to the dynamic response to intracavity power, increasing the final LED intensity also accelerates the response. 

\begin{figure}
	\includegraphics[width=\columnwidth]{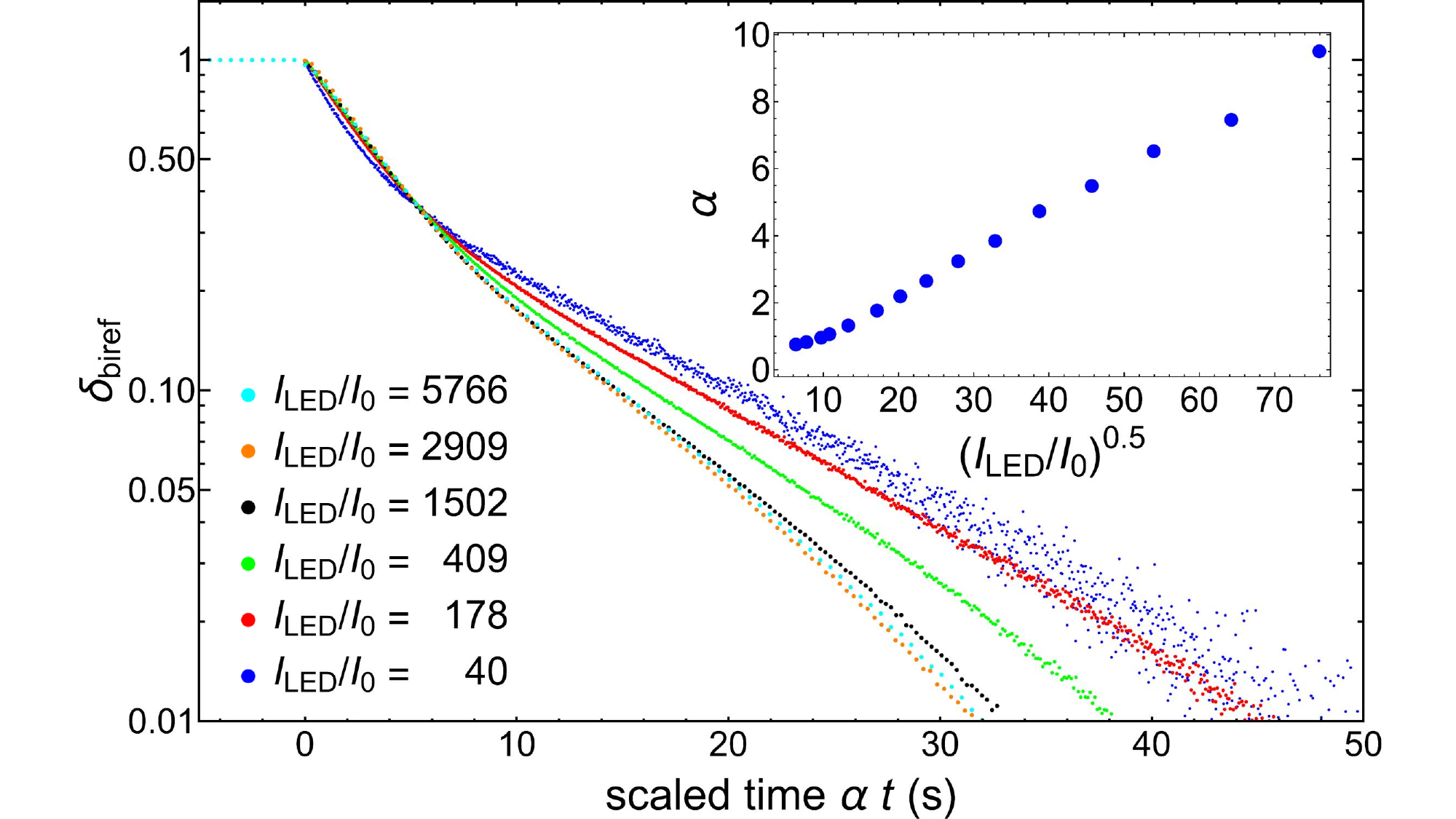}
	\caption{\label{Fig:TransLEDscale}
		Normalized transient response of $\delta_\mathrm{biref}$ to switching on a 535 nm LED to different final intensities 
		$I_\mathrm{LED}$ with $I_0 = 3.4~\upmu\mathrm{W/m^2}$.
		The intracavity power was fixed to $P_\mathrm{trans} = 12.5\,\upmu$W.
		The time axis is scaled by a factor $\alpha$ compared to the curve with $({I_\mathrm{LED}}/{I_{0}})^{0.5} = 10$ to match the initial part of the response for the different intensities.
		The inset shows the scaling factor $\alpha$ as a function of 
		$({I_\mathrm{LED}}/{I_{0}})^{0.5}$.
	}
\end{figure}

We overlap the curves at short times by stretching the time axis such that they agree at $\delta_\mathrm{biref} (\alpha\; t) = 1/e $, where  $\alpha = 1 $ refers to $({I_\mathrm{LED}}/{I_{0}})^{0.5}$ = 10. 
By scaling the time axis, as shown in Fig. \ref{Fig:TransLEDscale}, the initial decay of the curves can be overlapped.
The scaling factors $\alpha$ are approximately proportional to the square root of $({I_\mathrm{LED}}/{I_{0}})$ as shown in the inset.
At longer times a second exponential decay is visible. The time constant of the second exponential term decreases even faster with increasing LED intensity than the initial $1/e$ time constant.  


\section*{References}
\bibliography{PhotoBiref} 

\begin{thebibliography}{33}%
\makeatletter
\providecommand \@ifxundefined [1]{%
 \@ifx{#1\undefined}
}%
\providecommand \@ifnum [1]{%
 \ifnum #1\expandafter \@firstoftwo
 \else \expandafter \@secondoftwo
 \fi
}%
\providecommand \@ifx [1]{%
 \ifx #1\expandafter \@firstoftwo
 \else \expandafter \@secondoftwo
 \fi
}%
\providecommand \natexlab [1]{#1}%
\providecommand \enquote  [1]{``#1''}%
\providecommand \bibnamefont  [1]{#1}%
\providecommand \bibfnamefont [1]{#1}%
\providecommand \citenamefont [1]{#1}%
\providecommand \href@noop [0]{\@secondoftwo}%
\providecommand \href [0]{\begingroup \@sanitize@url \@href}%
\providecommand \@href[1]{\@@startlink{#1}\@@href}%
\providecommand \@@href[1]{\endgroup#1\@@endlink}%
\providecommand \@sanitize@url [0]{\catcode `\\12\catcode `\$12\catcode
  `\&12\catcode `\#12\catcode `\^12\catcode `\_12\catcode `\%12\relax}%
\providecommand \@@startlink[1]{}%
\providecommand \@@endlink[0]{}%
\providecommand \url  [0]{\begingroup\@sanitize@url \@url }%
\providecommand \@url [1]{\endgroup\@href {#1}{\urlprefix }}%
\providecommand \urlprefix  [0]{URL }%
\providecommand \Eprint [0]{\href }%
\providecommand \doibase [0]{https://doi.org/}%
\providecommand \selectlanguage [0]{\@gobble}%
\providecommand \bibinfo  [0]{\@secondoftwo}%
\providecommand \bibfield  [0]{\@secondoftwo}%
\providecommand \translation [1]{[#1]}%
\providecommand \BibitemOpen [0]{}%
\providecommand \bibitemStop [0]{}%
\providecommand \bibitemNoStop [0]{.\EOS\space}%
\providecommand \EOS [0]{\spacefactor3000\relax}%
\providecommand \BibitemShut  [1]{\csname bibitem#1\endcsname}%
\let\auto@bib@innerbib\@empty
\bibitem [{\citenamefont {Ludlow}\ \emph {et~al.}(2015)\citenamefont {Ludlow},
  \citenamefont {Boyd}, \citenamefont {Ye}, \citenamefont {Peik},\ and\
  \citenamefont {Schmidt}}]{lud15}%
  \BibitemOpen
  \bibfield  {author} {\bibinfo {author} {\bibfnamefont {A.~D.}\ \bibnamefont
  {Ludlow}}, \bibinfo {author} {\bibfnamefont {M.~M.}\ \bibnamefont {Boyd}},
  \bibinfo {author} {\bibfnamefont {J.}~\bibnamefont {Ye}}, \bibinfo {author}
  {\bibfnamefont {E.}~\bibnamefont {Peik}},\ and\ \bibinfo {author}
  {\bibfnamefont {P.~O.}\ \bibnamefont {Schmidt}},\ }\bibfield  {title}
  {\bibinfo {title} {Optical atomic clocks},\ }\href
  {https://doi.org/10.1103/RevModPhys.87.637} {\bibfield  {journal} {\bibinfo
  {journal} {Rev. Mod. Phys.}\ }\textbf {\bibinfo {volume} {87}},\ \bibinfo
  {pages} {637} (\bibinfo {year} {2015})}\BibitemShut {NoStop}%
\bibitem [{\citenamefont {Harry}\ \emph {et~al.}(2006)\citenamefont {Harry},
  \citenamefont {Armandula}, \citenamefont {Black}, \citenamefont {Crooks},
  \citenamefont {Cagnoli}, \citenamefont {Hough}, \citenamefont {Murray},
  \citenamefont {Reid}, \citenamefont {Rowan}, \citenamefont {Sneddon},
  \citenamefont {Fejer}, \citenamefont {Route},\ and\ \citenamefont
  {Penn}}]{har06b}%
  \BibitemOpen
  \bibfield  {author} {\bibinfo {author} {\bibfnamefont {G.~M.}\ \bibnamefont
  {Harry}}, \bibinfo {author} {\bibfnamefont {H.}~\bibnamefont {Armandula}},
  \bibinfo {author} {\bibfnamefont {E.}~\bibnamefont {Black}}, \bibinfo
  {author} {\bibfnamefont {D.~R.~M.}\ \bibnamefont {Crooks}}, \bibinfo {author}
  {\bibfnamefont {G.}~\bibnamefont {Cagnoli}}, \bibinfo {author} {\bibfnamefont
  {J.}~\bibnamefont {Hough}}, \bibinfo {author} {\bibfnamefont
  {P.}~\bibnamefont {Murray}}, \bibinfo {author} {\bibfnamefont
  {S.}~\bibnamefont {Reid}}, \bibinfo {author} {\bibfnamefont {S.}~\bibnamefont
  {Rowan}}, \bibinfo {author} {\bibfnamefont {P.}~\bibnamefont {Sneddon}},
  \bibinfo {author} {\bibfnamefont {M.~M.}\ \bibnamefont {Fejer}}, \bibinfo
  {author} {\bibfnamefont {R.}~\bibnamefont {Route}},\ and\ \bibinfo {author}
  {\bibfnamefont {S.~D.}\ \bibnamefont {Penn}},\ }\bibfield  {title} {\bibinfo
  {title} {Thermal noise from optical coatings in gravitational wave
  detectors},\ }\href {https://doi.org/10.1364/AO.45.001569} {\bibfield
  {journal} {\bibinfo  {journal} {Appl. Opt.}\ }\textbf {\bibinfo {volume}
  {45}},\ \bibinfo {pages} {1569} (\bibinfo {year} {2006})}\BibitemShut
  {NoStop}%
\bibitem [{\citenamefont {Granata}\ \emph {et~al.}(2020)\citenamefont
  {Granata}, \citenamefont {Amato}, \citenamefont {Cagnoli}, \citenamefont
  {Coulon}, \citenamefont {Degallaix}, \citenamefont {Forest}, \citenamefont
  {Mereni}, \citenamefont {Michel}, \citenamefont {Pinard}, \citenamefont
  {Sassolas},\ and\ \citenamefont {Teillon}}]{gra20a}%
  \BibitemOpen
  \bibfield  {author} {\bibinfo {author} {\bibfnamefont {M.}~\bibnamefont
  {Granata}}, \bibinfo {author} {\bibfnamefont {A.}~\bibnamefont {Amato}},
  \bibinfo {author} {\bibfnamefont {G.}~\bibnamefont {Cagnoli}}, \bibinfo
  {author} {\bibfnamefont {M.}~\bibnamefont {Coulon}}, \bibinfo {author}
  {\bibfnamefont {J.}~\bibnamefont {Degallaix}}, \bibinfo {author}
  {\bibfnamefont {D.}~\bibnamefont {Forest}}, \bibinfo {author} {\bibfnamefont
  {L.}~\bibnamefont {Mereni}}, \bibinfo {author} {\bibfnamefont
  {C.}~\bibnamefont {Michel}}, \bibinfo {author} {\bibfnamefont
  {L.}~\bibnamefont {Pinard}}, \bibinfo {author} {\bibfnamefont
  {B.}~\bibnamefont {Sassolas}},\ and\ \bibinfo {author} {\bibfnamefont
  {J.}~\bibnamefont {Teillon}},\ }\bibfield  {title} {\bibinfo {title}
  {Progress in the measurement and reduction of thermal noise in optical
  coatings for gravitational-wave detectors},\ }\href
  {https://doi.org/10.1364/AO.377293} {\bibfield  {journal} {\bibinfo
  {journal} {Appl. Opt.}\ }\textbf {\bibinfo {volume} {59}},\ \bibinfo {pages}
  {A229} (\bibinfo {year} {2020})}\BibitemShut {NoStop}%
\bibitem [{\citenamefont {Cole}\ \emph {et~al.}(2016)\citenamefont {Cole},
  \citenamefont {Zhang}, \citenamefont {Bjork}, \citenamefont {Follman},
  \citenamefont {Heu}, \citenamefont {Deutsch}, \citenamefont {Sonderhouse},
  \citenamefont {Robinson}, \citenamefont {Franz}, \citenamefont
  {Alexandrovski}, \citenamefont {Notcutt}, \citenamefont {Heckl},
  \citenamefont {Ye},\ and\ \citenamefont {Aspelmeyer}}]{col16}%
  \BibitemOpen
  \bibfield  {author} {\bibinfo {author} {\bibfnamefont {G.~D.}\ \bibnamefont
  {Cole}}, \bibinfo {author} {\bibfnamefont {W.}~\bibnamefont {Zhang}},
  \bibinfo {author} {\bibfnamefont {B.~J.}\ \bibnamefont {Bjork}}, \bibinfo
  {author} {\bibfnamefont {D.}~\bibnamefont {Follman}}, \bibinfo {author}
  {\bibfnamefont {P.}~\bibnamefont {Heu}}, \bibinfo {author} {\bibfnamefont
  {C.}~\bibnamefont {Deutsch}}, \bibinfo {author} {\bibfnamefont
  {L.}~\bibnamefont {Sonderhouse}}, \bibinfo {author} {\bibfnamefont
  {J.}~\bibnamefont {Robinson}}, \bibinfo {author} {\bibfnamefont
  {C.}~\bibnamefont {Franz}}, \bibinfo {author} {\bibfnamefont
  {A.}~\bibnamefont {Alexandrovski}}, \bibinfo {author} {\bibfnamefont
  {M.}~\bibnamefont {Notcutt}}, \bibinfo {author} {\bibfnamefont {O.~H.}\
  \bibnamefont {Heckl}}, \bibinfo {author} {\bibfnamefont {J.}~\bibnamefont
  {Ye}},\ and\ \bibinfo {author} {\bibfnamefont {M.}~\bibnamefont
  {Aspelmeyer}},\ }\bibfield  {title} {\bibinfo {title} {High-performance near-
  and mid-infrared crystalline coatings},\ }\href
  {https://doi.org/10.1364/OPTICA.3.000647} {\bibfield  {journal} {\bibinfo
  {journal} {Optica}\ }\textbf {\bibinfo {volume} {3}},\ \bibinfo {pages} {647}
  (\bibinfo {year} {2016})}\BibitemShut {NoStop}%
\bibitem [{\citenamefont {Yu}\ \emph {et~al.}(2023)\citenamefont {Yu},
  \citenamefont {H\"afner}, \citenamefont {Legero}, \citenamefont {Herbers},
  \citenamefont {Nicolodi}, \citenamefont {Ma}, \citenamefont {Riehle},
  \citenamefont {Sterr}, \citenamefont {Kedar}, \citenamefont {Robinson},
  \citenamefont {Oelker},\ and\ \citenamefont {Ye}}]{yu23a}%
  \BibitemOpen
  \bibfield  {author} {\bibinfo {author} {\bibfnamefont {J.}~\bibnamefont
  {Yu}}, \bibinfo {author} {\bibfnamefont {S.}~\bibnamefont {H\"afner}},
  \bibinfo {author} {\bibfnamefont {T.}~\bibnamefont {Legero}}, \bibinfo
  {author} {\bibfnamefont {S.}~\bibnamefont {Herbers}}, \bibinfo {author}
  {\bibfnamefont {D.}~\bibnamefont {Nicolodi}}, \bibinfo {author}
  {\bibfnamefont {C.~Y.}\ \bibnamefont {Ma}}, \bibinfo {author} {\bibfnamefont
  {F.}~\bibnamefont {Riehle}}, \bibinfo {author} {\bibfnamefont
  {U.}~\bibnamefont {Sterr}}, \bibinfo {author} {\bibfnamefont
  {D.}~\bibnamefont {Kedar}}, \bibinfo {author} {\bibfnamefont {J.~M.}\
  \bibnamefont {Robinson}}, \bibinfo {author} {\bibfnamefont {E.}~\bibnamefont
  {Oelker}},\ and\ \bibinfo {author} {\bibfnamefont {J.}~\bibnamefont {Ye}},\
  }\bibfield  {title} {\bibinfo {title} {Excess noise and photoinduced effects
  in highly reflective crystalline mirror coatings},\ }\href
  {https://doi.org/10.1103/PhysRevX.13.041002} {\bibfield  {journal} {\bibinfo
  {journal} {Phys. Rev. X}\ }\textbf {\bibinfo {volume} {13}},\ \bibinfo
  {pages} {041002} (\bibinfo {year} {2023})}\BibitemShut {NoStop}%
\bibitem [{\citenamefont {Kedar}\ \emph {et~al.}(2023)\citenamefont {Kedar},
  \citenamefont {Yu}, \citenamefont {Oelker}, \citenamefont {Staron},
  \citenamefont {Milner}, \citenamefont {Robinson}, \citenamefont {Legero},
  \citenamefont {Riehle}, \citenamefont {Sterr},\ and\ \citenamefont
  {Ye}}]{ked23}%
  \BibitemOpen
  \bibfield  {author} {\bibinfo {author} {\bibfnamefont {D.}~\bibnamefont
  {Kedar}}, \bibinfo {author} {\bibfnamefont {J.}~\bibnamefont {Yu}}, \bibinfo
  {author} {\bibfnamefont {E.}~\bibnamefont {Oelker}}, \bibinfo {author}
  {\bibfnamefont {A.}~\bibnamefont {Staron}}, \bibinfo {author} {\bibfnamefont
  {W.~R.}\ \bibnamefont {Milner}}, \bibinfo {author} {\bibfnamefont {J.~M.}\
  \bibnamefont {Robinson}}, \bibinfo {author} {\bibfnamefont {T.}~\bibnamefont
  {Legero}}, \bibinfo {author} {\bibfnamefont {F.}~\bibnamefont {Riehle}},
  \bibinfo {author} {\bibfnamefont {U.}~\bibnamefont {Sterr}},\ and\ \bibinfo
  {author} {\bibfnamefont {J.}~\bibnamefont {Ye}},\ }\bibfield  {title}
  {\bibinfo {title} {Frequency stability of cryogenic silicon cavities with
  semiconductor crystalline coatings},\ }\href
  {https://doi.org/10.1364/OPTICA.479462} {\bibfield  {journal} {\bibinfo
  {journal} {Optica}\ }\textbf {\bibinfo {volume} {10}},\ \bibinfo {pages}
  {464} (\bibinfo {year} {2023})}\BibitemShut {NoStop}%
\bibitem [{\citenamefont {Kraus}\ \emph {et~al.}(2025)\citenamefont {Kraus},
  \citenamefont {Herbers}, \citenamefont {Nauk}, \citenamefont {Sterr},
  \citenamefont {Lisdat},\ and\ \citenamefont {Schmidt}}]{kra25a}%
  \BibitemOpen
  \bibfield  {author} {\bibinfo {author} {\bibfnamefont {B.}~\bibnamefont
  {Kraus}}, \bibinfo {author} {\bibfnamefont {S.}~\bibnamefont {Herbers}},
  \bibinfo {author} {\bibfnamefont {C.}~\bibnamefont {Nauk}}, \bibinfo {author}
  {\bibfnamefont {U.}~\bibnamefont {Sterr}}, \bibinfo {author} {\bibfnamefont
  {C.}~\bibnamefont {Lisdat}},\ and\ \bibinfo {author} {\bibfnamefont {P.~O.}\
  \bibnamefont {Schmidt}},\ }\bibfield  {title} {\bibinfo {title} {Ultra-stable
  transportable ultraviolet clock laser using cancellation between
  photo-thermal and photo-birefringence noise},\ }\href
  {https://doi.org/10.1364/OL.544907} {\bibfield  {journal} {\bibinfo
  {journal} {Opt. Lett.}\ }\textbf {\bibinfo {volume} {50}},\ \bibinfo {pages}
  {658} (\bibinfo {year} {2025})}\BibitemShut {NoStop}%
\bibitem [{\citenamefont {Zhu}\ \emph {et~al.}(2023)\citenamefont {Zhu},
  \citenamefont {Cui}, \citenamefont {Kong}, \citenamefont {Yu}, \citenamefont
  {Jiang}, \citenamefont {Xu}, \citenamefont {Dai}, \citenamefont {Chen},\ and\
  \citenamefont {Pan}}]{zhu23}%
  \BibitemOpen
  \bibfield  {author} {\bibinfo {author} {\bibfnamefont {X.-Q.}\ \bibnamefont
  {Zhu}}, \bibinfo {author} {\bibfnamefont {X.-Y.}\ \bibnamefont {Cui}},
  \bibinfo {author} {\bibfnamefont {D.-Q.}\ \bibnamefont {Kong}}, \bibinfo
  {author} {\bibfnamefont {H.-W.}\ \bibnamefont {Yu}}, \bibinfo {author}
  {\bibfnamefont {X.}~\bibnamefont {Jiang}}, \bibinfo {author} {\bibfnamefont
  {P.}~\bibnamefont {Xu}}, \bibinfo {author} {\bibfnamefont {H.-N.}\
  \bibnamefont {Dai}}, \bibinfo {author} {\bibfnamefont {Y.-A.}\ \bibnamefont
  {Chen}},\ and\ \bibinfo {author} {\bibfnamefont {J.-W.}\ \bibnamefont
  {Pan}},\ }\bibfield  {title} {\bibinfo {title} {Photo-birefringent effects of
  crystalline coatings in ultra-stable cavities},\ }in\ \href
  {https://doi.org/10.1117/12.3007859} {\emph {\bibinfo {booktitle} {Fourteenth
  International Conference on Information Optics and Photonics (CIOP 2023)}}},\
  Vol.\ \bibinfo {volume} {12935},\ \bibinfo {editor} {edited by\ \bibinfo
  {editor} {\bibfnamefont {Y.}~\bibnamefont {Yang}}},\ \bibinfo {organization}
  {International Society for Optics and Photonics}\ (\bibinfo  {publisher}
  {SPIE},\ \bibinfo {year} {2023})\ p.\ \bibinfo {pages} {1293541}\BibitemShut
  {NoStop}%
\bibitem [{\citenamefont {Ma}\ \emph {et~al.}(2024)\citenamefont {Ma},
  \citenamefont {Yu}, \citenamefont {Legero}, \citenamefont {Herbers},
  \citenamefont {Nicolodi}, \citenamefont {Kempkes}, \citenamefont {Riehle},
  \citenamefont {Kedar}, \citenamefont {Robinson}, \citenamefont {Ye},\ and\
  \citenamefont {Sterr}}]{ma24a}%
  \BibitemOpen
  \bibfield  {author} {\bibinfo {author} {\bibfnamefont {C.~Y.}\ \bibnamefont
  {Ma}}, \bibinfo {author} {\bibfnamefont {J.}~\bibnamefont {Yu}}, \bibinfo
  {author} {\bibfnamefont {T.}~\bibnamefont {Legero}}, \bibinfo {author}
  {\bibfnamefont {S.}~\bibnamefont {Herbers}}, \bibinfo {author} {\bibfnamefont
  {D.}~\bibnamefont {Nicolodi}}, \bibinfo {author} {\bibfnamefont
  {M.}~\bibnamefont {Kempkes}}, \bibinfo {author} {\bibfnamefont
  {F.}~\bibnamefont {Riehle}}, \bibinfo {author} {\bibfnamefont
  {D.}~\bibnamefont {Kedar}}, \bibinfo {author} {\bibfnamefont {J.~M.}\
  \bibnamefont {Robinson}}, \bibinfo {author} {\bibfnamefont {J.}~\bibnamefont
  {Ye}},\ and\ \bibinfo {author} {\bibfnamefont {U.}~\bibnamefont {Sterr}},\
  }\bibfield  {title} {\bibinfo {title} {Ultrastable lasers: investigations of
  crystalline mirrors and closed cycle cooling at 124~{K}},\ }\href
  {https://doi.org/10.1088/1742-6596/2889/1/012055} {\bibfield  {journal}
  {\bibinfo  {journal} {J. Phys.: Conf. Ser.}\ }\textbf {\bibinfo {volume}
  {2889}},\ \bibinfo {pages} {012055} (\bibinfo {year} {2024})}\BibitemShut
  {NoStop}%
\bibitem [{\citenamefont {Wu}\ \emph {et~al.}(2025)\citenamefont {Wu},
  \citenamefont {Goswami}, \citenamefont {Tanioka},\ and\ \citenamefont
  {Ballmer}}]{wu25c}%
  \BibitemOpen
  \bibfield  {author} {\bibinfo {author} {\bibfnamefont {B.}~\bibnamefont
  {Wu}}, \bibinfo {author} {\bibfnamefont {S.}~\bibnamefont {Goswami}},
  \bibinfo {author} {\bibfnamefont {S.}~\bibnamefont {Tanioka}},\ and\ \bibinfo
  {author} {\bibfnamefont {S.}~\bibnamefont {Ballmer}},\ }\href
  {https://doi.org/10.48550/arXiv.2512.00594} {\bibinfo {title} {Birefringence
  of {AlGaAs}/{GaA}s coatings under above-band-gap illumination, {GR} noise and
  photo-optic transfer function}},\ \bibinfo {howpublished} {arXiv:2512.00594
  [physics.ins-det]} (\bibinfo {year} {2025}),\ \bibinfo {note} {lIGO Document
  P2500676-v2}\BibitemShut {NoStop}%
\bibitem [{\citenamefont {H{\"a}fner}\ \emph {et~al.}(2015)\citenamefont
  {H{\"a}fner}, \citenamefont {Falke}, \citenamefont {Grebing}, \citenamefont
  {Vogt}, \citenamefont {Legero}, \citenamefont {Merimaa}, \citenamefont
  {Lisdat},\ and\ \citenamefont {Sterr}}]{hae15a}%
  \BibitemOpen
  \bibfield  {author} {\bibinfo {author} {\bibfnamefont {S.}~\bibnamefont
  {H{\"a}fner}}, \bibinfo {author} {\bibfnamefont {S.}~\bibnamefont {Falke}},
  \bibinfo {author} {\bibfnamefont {C.}~\bibnamefont {Grebing}}, \bibinfo
  {author} {\bibfnamefont {S.}~\bibnamefont {Vogt}}, \bibinfo {author}
  {\bibfnamefont {T.}~\bibnamefont {Legero}}, \bibinfo {author} {\bibfnamefont
  {M.}~\bibnamefont {Merimaa}}, \bibinfo {author} {\bibfnamefont
  {C.}~\bibnamefont {Lisdat}},\ and\ \bibinfo {author} {\bibfnamefont
  {U.}~\bibnamefont {Sterr}},\ }\bibfield  {title} {\bibinfo {title} {$8 \times
  10^{-17}$ fractional laser frequency instability with a long room-temperature
  cavity},\ }\href {https://doi.org/10.1364/OL.40.002112} {\bibfield  {journal}
  {\bibinfo  {journal} {Opt. Lett.}\ }\textbf {\bibinfo {volume} {40}},\
  \bibinfo {pages} {2112} (\bibinfo {year} {2015})}\BibitemShut {NoStop}%
\bibitem [{\citenamefont {Cole}\ \emph {et~al.}(2023)\citenamefont {Cole},
  \citenamefont {Ballmer}, \citenamefont {Billingsley}, \citenamefont {Cata\~no
  Lopez}, \citenamefont {Fejer}, \citenamefont {Fritschel}, \citenamefont
  {Gretarsson}, \citenamefont {Harry}, \citenamefont {Kedar}, \citenamefont
  {Legero}, \citenamefont {Makarem}, \citenamefont {Penn}, \citenamefont
  {Reitze}, \citenamefont {Steinlechner}, \citenamefont {Sterr}, \citenamefont
  {Tanioka}, \citenamefont {Truong}, \citenamefont {Ye},\ and\ \citenamefont
  {Yu}}]{col23}%
  \BibitemOpen
  \bibfield  {author} {\bibinfo {author} {\bibfnamefont {G.~D.}\ \bibnamefont
  {Cole}}, \bibinfo {author} {\bibfnamefont {S.}~\bibnamefont {Ballmer}},
  \bibinfo {author} {\bibfnamefont {G.}~\bibnamefont {Billingsley}}, \bibinfo
  {author} {\bibfnamefont {S.~B.}\ \bibnamefont {Cata\~no Lopez}}, \bibinfo
  {author} {\bibfnamefont {M.}~\bibnamefont {Fejer}}, \bibinfo {author}
  {\bibfnamefont {P.}~\bibnamefont {Fritschel}}, \bibinfo {author}
  {\bibfnamefont {A.~M.}\ \bibnamefont {Gretarsson}}, \bibinfo {author}
  {\bibfnamefont {G.~M.}\ \bibnamefont {Harry}}, \bibinfo {author}
  {\bibfnamefont {D.}~\bibnamefont {Kedar}}, \bibinfo {author} {\bibfnamefont
  {T.}~\bibnamefont {Legero}}, \bibinfo {author} {\bibfnamefont
  {C.}~\bibnamefont {Makarem}}, \bibinfo {author} {\bibfnamefont {S.~D.}\
  \bibnamefont {Penn}}, \bibinfo {author} {\bibfnamefont {D.}~\bibnamefont
  {Reitze}}, \bibinfo {author} {\bibfnamefont {J.}~\bibnamefont
  {Steinlechner}}, \bibinfo {author} {\bibfnamefont {U.}~\bibnamefont {Sterr}},
  \bibinfo {author} {\bibfnamefont {S.}~\bibnamefont {Tanioka}}, \bibinfo
  {author} {\bibfnamefont {G.~W.}\ \bibnamefont {Truong}}, \bibinfo {author}
  {\bibfnamefont {J.}~\bibnamefont {Ye}},\ and\ \bibinfo {author}
  {\bibfnamefont {J.}~\bibnamefont {Yu}},\ }\bibfield  {title} {\bibinfo
  {title} {Substrate-transferred {GaAs}/{AlGaAs} crystalline coatings for
  gravitational-wave detectors},\ }\href {https://doi.org/10.1063/5.0140663}
  {\bibfield  {journal} {\bibinfo  {journal} {Appl. Phys. Lett.}\ }\textbf
  {\bibinfo {volume} {122}},\ \bibinfo {pages} {110502} (\bibinfo {year}
  {2023})}\BibitemShut {NoStop}%
\bibitem [{\citenamefont {Legero}\ \emph {et~al.}(2010)\citenamefont {Legero},
  \citenamefont {Kessler},\ and\ \citenamefont {Sterr}}]{leg10}%
  \BibitemOpen
  \bibfield  {author} {\bibinfo {author} {\bibfnamefont {T.}~\bibnamefont
  {Legero}}, \bibinfo {author} {\bibfnamefont {T.}~\bibnamefont {Kessler}},\
  and\ \bibinfo {author} {\bibfnamefont {U.}~\bibnamefont {Sterr}},\ }\bibfield
   {title} {\bibinfo {title} {Tuning the thermal expansion properties of
  optical reference cavities with fused silica mirrors},\ }\href
  {https://doi.org/10.1364/JOSAB.27.000914} {\bibfield  {journal} {\bibinfo
  {journal} {J. Opt. Soc. Am. B}\ }\textbf {\bibinfo {volume} {27}},\ \bibinfo
  {pages} {914} (\bibinfo {year} {2010})}\BibitemShut {NoStop}%
\bibitem [{\citenamefont {Kraus}(2023)}]{kra23a}%
  \BibitemOpen
  \bibfield  {author} {\bibinfo {author} {\bibfnamefont {B.}~\bibnamefont
  {Kraus}},\ }\emph {\bibinfo {title} {A highly stable {UV} clock laser}},\
  \href {https://doi.org/10.15488/15360} {Ph.D. thesis},\ \bibinfo  {school}
  {QUEST-Leibniz-Forschungsschule der Gottfried Wilhelm Leibniz Universit\"at
  Hannover} (\bibinfo {year} {2023})\BibitemShut {NoStop}%
\bibitem [{\citenamefont {Yu}(2023)}]{yu23}%
  \BibitemOpen
  \bibfield  {author} {\bibinfo {author} {\bibfnamefont {J.}~\bibnamefont
  {Yu}},\ }\emph {\bibinfo {title} {Cryogenic silicon {Fabry}-{Perot} resonator
  with {Al$_{0.92}$Ga$_{0.08}$As/GaAs} mirror coatings.}},\ \href
  {https://doi.org/10.15488/13416} {Ph.D. thesis},\ \bibinfo  {school}
  {QUEST-Leibniz-Forschungsschule der Gottfried Wilhelm Leibniz Universit\"at
  Hannover} (\bibinfo {year} {2023})\BibitemShut {NoStop}%
\bibitem [{\citenamefont {Matei}\ \emph {et~al.}(2017)\citenamefont {Matei},
  \citenamefont {Legero}, \citenamefont {H\"afner}, \citenamefont {Grebing},
  \citenamefont {Weyrich}, \citenamefont {Zhang}, \citenamefont {Sonderhouse},
  \citenamefont {Robinson}, \citenamefont {Ye}, \citenamefont {Riehle},\ and\
  \citenamefont {Sterr}}]{mat17a}%
  \BibitemOpen
  \bibfield  {author} {\bibinfo {author} {\bibfnamefont {D.~G.}\ \bibnamefont
  {Matei}}, \bibinfo {author} {\bibfnamefont {T.}~\bibnamefont {Legero}},
  \bibinfo {author} {\bibfnamefont {S.}~\bibnamefont {H\"afner}}, \bibinfo
  {author} {\bibfnamefont {C.}~\bibnamefont {Grebing}}, \bibinfo {author}
  {\bibfnamefont {R.}~\bibnamefont {Weyrich}}, \bibinfo {author} {\bibfnamefont
  {W.}~\bibnamefont {Zhang}}, \bibinfo {author} {\bibfnamefont
  {L.}~\bibnamefont {Sonderhouse}}, \bibinfo {author} {\bibfnamefont {J.~M.}\
  \bibnamefont {Robinson}}, \bibinfo {author} {\bibfnamefont {J.}~\bibnamefont
  {Ye}}, \bibinfo {author} {\bibfnamefont {F.}~\bibnamefont {Riehle}},\ and\
  \bibinfo {author} {\bibfnamefont {U.}~\bibnamefont {Sterr}},\ }\bibfield
  {title} {\bibinfo {title} {$1.5~\mu$m lasers with sub-{10 mHz} linewidth},\
  }\href {https://doi.org/10.1103/PhysRevLett.118.263202} {\bibfield  {journal}
  {\bibinfo  {journal} {Phys. Rev. Lett.}\ }\textbf {\bibinfo {volume} {118}},\
  \bibinfo {pages} {263202} (\bibinfo {year} {2017})}\BibitemShut {NoStop}%
\bibitem [{\citenamefont {Penzkofer}\ and\ \citenamefont
  {Bugayev}(1989)}]{pen89}%
  \BibitemOpen
  \bibfield  {author} {\bibinfo {author} {\bibfnamefont {A.}~\bibnamefont
  {Penzkofer}}\ and\ \bibinfo {author} {\bibfnamefont {A.~A.}\ \bibnamefont
  {Bugayev}},\ }\bibfield  {title} {\bibinfo {title} {Two-photon absorption and
  emission dynamics of bulk {GaAs}},\ }\href
  {https://doi.org/10.1007/BF02027300} {\bibfield  {journal} {\bibinfo
  {journal} {Opt. Quantum Electron.}\ }\textbf {\bibinfo {volume} {21}},\
  \bibinfo {pages} {283} (\bibinfo {year} {1989})}\BibitemShut {NoStop}%
\bibitem [{\citenamefont {Hurlbut}\ \emph {et~al.}(2007)\citenamefont
  {Hurlbut}, \citenamefont {Lee}, \citenamefont {Vodopyanov}, \citenamefont
  {Kuo},\ and\ \citenamefont {Fejer}}]{hur07}%
  \BibitemOpen
  \bibfield  {author} {\bibinfo {author} {\bibfnamefont {W.~C.}\ \bibnamefont
  {Hurlbut}}, \bibinfo {author} {\bibfnamefont {Y.-S.}\ \bibnamefont {Lee}},
  \bibinfo {author} {\bibfnamefont {K.~L.}\ \bibnamefont {Vodopyanov}},
  \bibinfo {author} {\bibfnamefont {P.~S.}\ \bibnamefont {Kuo}},\ and\ \bibinfo
  {author} {\bibfnamefont {M.~M.}\ \bibnamefont {Fejer}},\ }\bibfield  {title}
  {\bibinfo {title} {Multiphoton absorption and nonlinear refraction of {GaAs}
  in the mid-infrared},\ }\href {https://doi.org/10.1364/OL.32.000668}
  {\bibfield  {journal} {\bibinfo  {journal} {Opt. Lett.}\ }\textbf {\bibinfo
  {volume} {32}},\ \bibinfo {pages} {668} (\bibinfo {year} {2007})}\BibitemShut
  {NoStop}%
\bibitem [{\citenamefont {Adachi}(1992)}]{ada92}%
  \BibitemOpen
  \bibfield  {author} {\bibinfo {author} {\bibfnamefont {S.}~\bibnamefont
  {Adachi}},\ }\bibinfo {title} {Elastooptic and electrooptic effects},\ in\
  \href {https://doi.org/10.1002/352760281X.ch9} {\emph {\bibinfo {booktitle}
  {Physical Properties of {III}-{V} Semiconductor Compounds}}}\ (\bibinfo
  {publisher} {John Wiley \& Sons, Ltd},\ \bibinfo {year} {1992})\
  Chap.~\bibinfo {chapter} {9}, pp.\ \bibinfo {pages} {193--222}\BibitemShut
  {NoStop}%
\bibitem [{\citenamefont {Adachi}\ and\ \citenamefont {Oe}(1983)}]{ada83}%
  \BibitemOpen
  \bibfield  {author} {\bibinfo {author} {\bibfnamefont {S.}~\bibnamefont
  {Adachi}}\ and\ \bibinfo {author} {\bibfnamefont {K.}~\bibnamefont {Oe}},\
  }\bibfield  {title} {\bibinfo {title} {Internal strain and photoelastic
  effects in {Ga$_{1-x}$Al$_{x}$As/GaAs} and
  {In$_{1-x}$Ga$_x$As$_y$P$_{1-y}$/InP} crystals},\ }\href
  {https://doi.org/10.1063/1.331898} {\bibfield  {journal} {\bibinfo  {journal}
  {J. Appl. Phys.}\ }\textbf {\bibinfo {volume} {54}},\ \bibinfo {pages} {6620}
  (\bibinfo {year} {1983})}\BibitemShut {NoStop}%
\bibitem [{\citenamefont {Winkler}\ \emph {et~al.}(2021)\citenamefont
  {Winkler}, \citenamefont {Perner}, \citenamefont {Truong}, \citenamefont
  {Zhao}, \citenamefont {Bachmann}, \citenamefont {Mayer}, \citenamefont
  {Fellinger}, \citenamefont {Follman}, \citenamefont {Heu}, \citenamefont
  {Deutsch}, \citenamefont {Bailey}, \citenamefont {Peelaers}, \citenamefont
  {Puchegger}, \citenamefont {Fleisher}, \citenamefont {Cole},\ and\
  \citenamefont {Heckl}}]{win21}%
  \BibitemOpen
  \bibfield  {author} {\bibinfo {author} {\bibfnamefont {G.}~\bibnamefont
  {Winkler}}, \bibinfo {author} {\bibfnamefont {L.~W.}\ \bibnamefont {Perner}},
  \bibinfo {author} {\bibfnamefont {G.-W.}\ \bibnamefont {Truong}}, \bibinfo
  {author} {\bibfnamefont {G.}~\bibnamefont {Zhao}}, \bibinfo {author}
  {\bibfnamefont {D.}~\bibnamefont {Bachmann}}, \bibinfo {author}
  {\bibfnamefont {A.~S.}\ \bibnamefont {Mayer}}, \bibinfo {author}
  {\bibfnamefont {J.}~\bibnamefont {Fellinger}}, \bibinfo {author}
  {\bibfnamefont {D.}~\bibnamefont {Follman}}, \bibinfo {author} {\bibfnamefont
  {P.}~\bibnamefont {Heu}}, \bibinfo {author} {\bibfnamefont {C.}~\bibnamefont
  {Deutsch}}, \bibinfo {author} {\bibfnamefont {D.~M.}\ \bibnamefont {Bailey}},
  \bibinfo {author} {\bibfnamefont {H.}~\bibnamefont {Peelaers}}, \bibinfo
  {author} {\bibfnamefont {S.}~\bibnamefont {Puchegger}}, \bibinfo {author}
  {\bibfnamefont {A.~J.}\ \bibnamefont {Fleisher}}, \bibinfo {author}
  {\bibfnamefont {G.~D.}\ \bibnamefont {Cole}},\ and\ \bibinfo {author}
  {\bibfnamefont {O.~H.}\ \bibnamefont {Heckl}},\ }\bibfield  {title} {\bibinfo
  {title} {Mid-infrared interference coatings with excess optical loss below
  10~ppm},\ }\href {https://doi.org/10.1364/OPTICA.405938} {\bibfield
  {journal} {\bibinfo  {journal} {Optica}\ }\textbf {\bibinfo {volume} {8}},\
  \bibinfo {pages} {686} (\bibinfo {year} {2021})},\ \bibinfo {note} {see
  erratum: https://doi.org/10.1364/OPTICA.520398 \cite{per24}}\BibitemShut
  {NoStop}%
\bibitem [{\citenamefont {Kushida}\ \emph {et~al.}(2017)\citenamefont
  {Kushida}, \citenamefont {Ohmori},\ and\ \citenamefont {Sakaki}}]{kus17}%
  \BibitemOpen
  \bibfield  {author} {\bibinfo {author} {\bibfnamefont {T.}~\bibnamefont
  {Kushida}}, \bibinfo {author} {\bibfnamefont {M.}~\bibnamefont {Ohmori}},\
  and\ \bibinfo {author} {\bibfnamefont {H.}~\bibnamefont {Sakaki}},\
  }\bibfield  {title} {\bibinfo {title} {Photocurrent and photoluminescence
  characteristics of {AlGaAs}/{GaAs} double-heterostructures with a pair of
  two-dimensional electron and hole channels},\ }\href
  {https://doi.org/10.1063/1.5001507} {\bibfield  {journal} {\bibinfo
  {journal} {J. Appl. Phys.}\ }\textbf {\bibinfo {volume} {122}},\ \bibinfo
  {pages} {104502} (\bibinfo {year} {2017})}\BibitemShut {NoStop}%
\bibitem [{\citenamefont {Shockley}(1949)}]{sho49}%
  \BibitemOpen
  \bibfield  {author} {\bibinfo {author} {\bibfnamefont {W.}~\bibnamefont
  {Shockley}},\ }\bibfield  {title} {\bibinfo {title} {The theory of p-n
  junctions in semiconductors and p-n junction transistors},\ }\href
  {https://doi.org/10.1002/j.1538-7305.1949.tb03645.x} {\bibfield  {journal}
  {\bibinfo  {journal} {Bell Sys. Tech. J.}\ }\textbf {\bibinfo {volume}
  {28}},\ \bibinfo {pages} {435} (\bibinfo {year} {1949})}\BibitemShut
  {NoStop}%
\bibitem [{\citenamefont {Fujita}\ \emph {et~al.}(2021)\citenamefont {Fujita},
  \citenamefont {Hayashi}, \citenamefont {Kohda}, \citenamefont {Ritzmann},
  \citenamefont {Ludwig}, \citenamefont {Nitta}, \citenamefont {Wieck},\ and\
  \citenamefont {Oiwa}}]{fuj21}%
  \BibitemOpen
  \bibfield  {author} {\bibinfo {author} {\bibfnamefont {T.}~\bibnamefont
  {Fujita}}, \bibinfo {author} {\bibfnamefont {R.}~\bibnamefont {Hayashi}},
  \bibinfo {author} {\bibfnamefont {M.}~\bibnamefont {Kohda}}, \bibinfo
  {author} {\bibfnamefont {J.}~\bibnamefont {Ritzmann}}, \bibinfo {author}
  {\bibfnamefont {A.}~\bibnamefont {Ludwig}}, \bibinfo {author} {\bibfnamefont
  {J.}~\bibnamefont {Nitta}}, \bibinfo {author} {\bibfnamefont {A.~D.}\
  \bibnamefont {Wieck}},\ and\ \bibinfo {author} {\bibfnamefont
  {A.}~\bibnamefont {Oiwa}},\ }\bibfield  {title} {\bibinfo {title}
  {Distinguishing persistent effects in an undoped {GaAs/AlGaAs} quantum well
  by top-gate-dependent illumination},\ }\href
  {https://doi.org/10.1063/5.0047558} {\bibfield  {journal} {\bibinfo
  {journal} {J. Appl. Phys.}\ }\textbf {\bibinfo {volume} {129}},\ \bibinfo
  {pages} {234301} (\bibinfo {year} {2021})}\BibitemShut {NoStop}%
\bibitem [{\citenamefont {Nathan}(1986)}]{nat86}%
  \BibitemOpen
  \bibfield  {author} {\bibinfo {author} {\bibfnamefont {M.~I.}\ \bibnamefont
  {Nathan}},\ }\bibfield  {title} {\bibinfo {title} {Persistent
  photoconductivity in {AlGaAs}/{GaAs} modulation doped layers and field effect
  transistors: {A} review},\ }\href
  {https://doi.org/10.1016/0038-1101(86)90035-3} {\bibfield  {journal}
  {\bibinfo  {journal} {Solid-State Electron.}\ }\textbf {\bibinfo {volume}
  {29}},\ \bibinfo {pages} {167} (\bibinfo {year} {1986})}\BibitemShut
  {NoStop}%
\bibitem [{\citenamefont {Lin}\ \emph {et~al.}(2017)\citenamefont {Lin},
  \citenamefont {Shuvra}, \citenamefont {McNamara}, \citenamefont {Gong},
  \citenamefont {Liao}, \citenamefont {Davidson}, \citenamefont {Walsh},
  \citenamefont {Alles},\ and\ \citenamefont {Alphenaar}}]{lin17}%
  \BibitemOpen
  \bibfield  {author} {\bibinfo {author} {\bibfnamefont {J.~T.}\ \bibnamefont
  {Lin}}, \bibinfo {author} {\bibfnamefont {P.~D.}\ \bibnamefont {Shuvra}},
  \bibinfo {author} {\bibfnamefont {S.}~\bibnamefont {McNamara}}, \bibinfo
  {author} {\bibfnamefont {H.}~\bibnamefont {Gong}}, \bibinfo {author}
  {\bibfnamefont {W.}~\bibnamefont {Liao}}, \bibinfo {author} {\bibfnamefont
  {J.~L.}\ \bibnamefont {Davidson}}, \bibinfo {author} {\bibfnamefont {K.~M.}\
  \bibnamefont {Walsh}}, \bibinfo {author} {\bibfnamefont {M.~L.}\ \bibnamefont
  {Alles}},\ and\ \bibinfo {author} {\bibfnamefont {B.~W.}\ \bibnamefont
  {Alphenaar}},\ }\bibfield  {title} {\bibinfo {title} {Near-surface electronic
  contribution to semiconductor elasticity},\ }\href
  {https://doi.org/10.1103/PhysRevApplied.8.034013} {\bibfield  {journal}
  {\bibinfo  {journal} {Phys. Rev. Appl.}\ }\textbf {\bibinfo {volume} {8}},\
  \bibinfo {pages} {034013} (\bibinfo {year} {2017})}\BibitemShut {NoStop}%
\bibitem [{\citenamefont {Ghosh}(2004)}]{gho04}%
  \BibitemOpen
  \bibfield  {author} {\bibinfo {author} {\bibfnamefont {S.}~\bibnamefont
  {Ghosh}},\ }\bibfield  {title} {\bibinfo {title} {Slow relaxation of
  nonequilibrated photo-carriers in semiconductors},\ }\href
  {https://doi.org/10.1080/01411590410001690909} {\bibfield  {journal}
  {\bibinfo  {journal} {Phase Transitions}\ }\textbf {\bibinfo {volume} {77}},\
  \bibinfo {pages} {791} (\bibinfo {year} {2004})}\BibitemShut {NoStop}%
\bibitem [{\citenamefont {Aspnes}\ \emph {et~al.}(1986)\citenamefont {Aspnes},
  \citenamefont {Kelso}, \citenamefont {Logan},\ and\ \citenamefont
  {Bhat}}]{asp86a}%
  \BibitemOpen
  \bibfield  {author} {\bibinfo {author} {\bibfnamefont {D.~E.}\ \bibnamefont
  {Aspnes}}, \bibinfo {author} {\bibfnamefont {S.~M.}\ \bibnamefont {Kelso}},
  \bibinfo {author} {\bibfnamefont {R.~A.}\ \bibnamefont {Logan}},\ and\
  \bibinfo {author} {\bibfnamefont {R.}~\bibnamefont {Bhat}},\ }\bibfield
  {title} {\bibinfo {title} {Optical properties of {Al$_x$Ga$_{1-x}$As}},\
  }\href {https://doi.org/10.1063/1.337426} {\bibfield  {journal} {\bibinfo
  {journal} {J. Appl. Phys.}\ }\textbf {\bibinfo {volume} {60}},\ \bibinfo
  {pages} {754} (\bibinfo {year} {1986})}\BibitemShut {NoStop}%
\bibitem [{\citenamefont {Farsi}\ \emph {et~al.}(2012)\citenamefont {Farsi},
  \citenamefont {Siciliani~de Cumis}, \citenamefont {Marino},\ and\
  \citenamefont {Marin}}]{far12}%
  \BibitemOpen
  \bibfield  {author} {\bibinfo {author} {\bibfnamefont {A.}~\bibnamefont
  {Farsi}}, \bibinfo {author} {\bibfnamefont {M.}~\bibnamefont {Siciliani~de
  Cumis}}, \bibinfo {author} {\bibfnamefont {F.}~\bibnamefont {Marino}},\ and\
  \bibinfo {author} {\bibfnamefont {F.}~\bibnamefont {Marin}},\ }\bibfield
  {title} {\bibinfo {title} {Photothermal and thermo-refractive effects in high
  reflectivity mirrors at room and cryogenic temperature},\ }\href
  {https://doi.org/10.1063/1.3684626} {\bibfield  {journal} {\bibinfo
  {journal} {J. Appl. Phys.}\ }\textbf {\bibinfo {volume} {111}},\ \bibinfo
  {pages} {043101} (\bibinfo {year} {2012})}\BibitemShut {NoStop}%
\bibitem [{\citenamefont {Herbers}(2021)}]{her21_short}%
  \BibitemOpen
  \bibfield  {author} {\bibinfo {author} {\bibfnamefont {S.}~\bibnamefont
  {Herbers}},\ }\emph {\bibinfo {title} {Transportable ultra-stable laser
  system with an instability down to $10^{-16}$}},\ \href
  {https://doi.org/10.15488/11624} {Ph.D. thesis},\ \bibinfo  {school}
  {QUEST-Leibniz-Forschungsschule der Gottfried Wilhelm Leibniz Universit\"at
  Hannover} (\bibinfo {year} {2021})\BibitemShut {NoStop}%
\bibitem [{\citenamefont {De~Rosa}\ \emph {et~al.}(2002)\citenamefont
  {De~Rosa}, \citenamefont {Conti}, \citenamefont {Cerdonio}, \citenamefont
  {Pinard},\ and\ \citenamefont {Marin}}]{ros02}%
  \BibitemOpen
  \bibfield  {author} {\bibinfo {author} {\bibfnamefont {M.}~\bibnamefont
  {De~Rosa}}, \bibinfo {author} {\bibfnamefont {L.}~\bibnamefont {Conti}},
  \bibinfo {author} {\bibfnamefont {M.}~\bibnamefont {Cerdonio}}, \bibinfo
  {author} {\bibfnamefont {M.}~\bibnamefont {Pinard}},\ and\ \bibinfo {author}
  {\bibfnamefont {F.}~\bibnamefont {Marin}},\ }\bibfield  {title} {\bibinfo
  {title} {Experimental measurement of the dynamic photothermal effect in
  {Fabry}-{Perot} cavities for gravitational wave detectors},\ }\href
  {https://doi.org/10.1103/PhysRevLett.89.237402} {\bibfield  {journal}
  {\bibinfo  {journal} {Phys. Rev. Lett.}\ }\textbf {\bibinfo {volume} {89}},\
  \bibinfo {pages} {237402} (\bibinfo {year} {2002})}\BibitemShut {NoStop}%
\bibitem [{\citenamefont {Zhu}\ \emph {et~al.}(2024)\citenamefont {Zhu},
  \citenamefont {Cui}, \citenamefont {Kong}, \citenamefont {Yu}, \citenamefont
  {Zhai}, \citenamefont {Zheng}, \citenamefont {Xie}, \citenamefont {Zhang},
  \citenamefont {Jiang}, \citenamefont {Zhang}, \citenamefont {Xu},
  \citenamefont {Dai}, \citenamefont {Chen},\ and\ \citenamefont
  {Pan}}]{zhu24}%
  \BibitemOpen
  \bibfield  {author} {\bibinfo {author} {\bibfnamefont {X.-Q.}\ \bibnamefont
  {Zhu}}, \bibinfo {author} {\bibfnamefont {X.-Y.}\ \bibnamefont {Cui}},
  \bibinfo {author} {\bibfnamefont {D.-Q.}\ \bibnamefont {Kong}}, \bibinfo
  {author} {\bibfnamefont {H.-W.}\ \bibnamefont {Yu}}, \bibinfo {author}
  {\bibfnamefont {X.-M.}\ \bibnamefont {Zhai}}, \bibinfo {author}
  {\bibfnamefont {M.-Y.}\ \bibnamefont {Zheng}}, \bibinfo {author}
  {\bibfnamefont {X.-P.}\ \bibnamefont {Xie}}, \bibinfo {author} {\bibfnamefont
  {Q.}~\bibnamefont {Zhang}}, \bibinfo {author} {\bibfnamefont
  {X.}~\bibnamefont {Jiang}}, \bibinfo {author} {\bibfnamefont {X.-B.}\
  \bibnamefont {Zhang}}, \bibinfo {author} {\bibfnamefont {P.}~\bibnamefont
  {Xu}}, \bibinfo {author} {\bibfnamefont {H.-N.}\ \bibnamefont {Dai}},
  \bibinfo {author} {\bibfnamefont {Y.-A.}\ \bibnamefont {Chen}},\ and\
  \bibinfo {author} {\bibfnamefont {J.-W.}\ \bibnamefont {Pan}},\ }\bibfield
  {title} {\bibinfo {title} {An ultrastable 1397-nm laser stabilized by a
  crystalline-coated room-temperature cavity},\ }\href
  {https://doi.org/10.1063/5.0200553} {\bibfield  {journal} {\bibinfo
  {journal} {Rev. Sci. Instrum.}\ }\textbf {\bibinfo {volume} {95}},\ \bibinfo
  {pages} {083002} (\bibinfo {year} {2024})}\BibitemShut {NoStop}%
\bibitem [{\citenamefont {Perner}\ \emph {et~al.}(2024)\citenamefont {Perner},
  \citenamefont {Winkler}, \citenamefont {Truong}, \citenamefont {Zhao},
  \citenamefont {Bachmann}, \citenamefont {Mayer}, \citenamefont {Fellinger},
  \citenamefont {Follman}, \citenamefont {Heu}, \citenamefont {Deutsch},
  \citenamefont {Bailey}, \citenamefont {Peelaers}, \citenamefont {Puchegger},
  \citenamefont {Fleisher}, \citenamefont {Cole},\ and\ \citenamefont
  {Heckl}}]{per24}%
  \BibitemOpen
  \bibfield  {author} {\bibinfo {author} {\bibfnamefont {L.~W.}\ \bibnamefont
  {Perner}}, \bibinfo {author} {\bibfnamefont {G.}~\bibnamefont {Winkler}},
  \bibinfo {author} {\bibfnamefont {G.-W.}\ \bibnamefont {Truong}}, \bibinfo
  {author} {\bibfnamefont {G.}~\bibnamefont {Zhao}}, \bibinfo {author}
  {\bibfnamefont {D.}~\bibnamefont {Bachmann}}, \bibinfo {author}
  {\bibfnamefont {A.~S.}\ \bibnamefont {Mayer}}, \bibinfo {author}
  {\bibfnamefont {J.}~\bibnamefont {Fellinger}}, \bibinfo {author}
  {\bibfnamefont {D.}~\bibnamefont {Follman}}, \bibinfo {author} {\bibfnamefont
  {P.}~\bibnamefont {Heu}}, \bibinfo {author} {\bibfnamefont {C.}~\bibnamefont
  {Deutsch}}, \bibinfo {author} {\bibfnamefont {D.~M.}\ \bibnamefont {Bailey}},
  \bibinfo {author} {\bibfnamefont {H.}~\bibnamefont {Peelaers}}, \bibinfo
  {author} {\bibfnamefont {S.}~\bibnamefont {Puchegger}}, \bibinfo {author}
  {\bibfnamefont {A.~J.}\ \bibnamefont {Fleisher}}, \bibinfo {author}
  {\bibfnamefont {G.~D.}\ \bibnamefont {Cole}},\ and\ \bibinfo {author}
  {\bibfnamefont {O.~H.}\ \bibnamefont {Heckl}},\ }\bibfield  {title} {\bibinfo
  {title} {Mid-infrared interference coatings with excess optical loss below
  10~ppm: erratum},\ }\href {https://doi.org/10.1364/OPTICA.520398} {\bibfield
  {journal} {\bibinfo  {journal} {Optica}\ }\textbf {\bibinfo {volume} {11}},\
  \bibinfo {pages} {619} (\bibinfo {year} {2024})}\BibitemShut {NoStop}%
\end{thebibliography}%
\clearpage
\appendix

\end{document}